\documentstyle[prl,aps,floats,epsf,epsfig]{revtex}

\begin{document}

\draft

\title{Commensurate and Incommensurate $O(n)$ Spin Systems: 
Novel Even-Odd Effects, A Generalized 
Mermin-Wagner-Coleman Theorem, and
Ground States}

\author{Zohar Nussinov*}
\address{Institute Lorentz for Theoretical Physics, Leiden University\\
P.O.B. 9506, 2300 RA Leiden, The Netherlands}
\date{\today}

\twocolumn[

\widetext
\begin{@twocolumnfalse}

\maketitle

\begin{abstract}

We examine $n$ component spin
systems with arbitrary two 
spin interactions (of unspecified
range) within a 
general framework to highlight some
new subtleties present in incommensurate systems.
We determine the 
ground states of all translationally
invariant 
$O(n \ge 2)$ systems and
prove that barring commensurability effects 
they are always spiral- no other ground
states are possible.
We study the effect of thermal 
fluctuations on the ground states
to discover a novel odd-even $n$ 
effect. Soft spin analysis suggests
that algebraic long range order
is possible in certain
frustrated incommensurate even
$n$ systems while their odd $n$
counterparts exhibit an exponential decay of
correlations. We illustrate
that many frustrated incommensurate continuous
spin systems display smectic like thermodynamics. 
We report on a generalized 
Mermin-Wagner-Coleman theorem
for all two dimensional systems 
(of arbitrary range) 
with analytic kernels
in momentum space. A new relation
between generalization Mermin-Wagner-Coleman
bounds and dynamics is further reported. 
We suggest a link between a
generalized Mermin-Wagner-Coleman
theorem to divergent decoherence
(or bandwidth) time scale in the quantum context.
A generalization of the Peierls bound for commensurate
systems with long range interactions is also discussed.
We conclude with a 
discussion of $O(n)$ 
spin dynamics in the 
general case.

\end{abstract}

\vspace{0.5cm}

\narrowtext

\end{@twocolumnfalse}
]

\section{Introduction}

In this article (constituting a portion
of my thesis \cite{thesis}), we aim to unveil some of general
properties of $O(n)$ spin systems having 
two-spin interactions.  We investigate
Ising and $O(n)$ ground
states, derive generalized 
Mermin-Wagner inequalities
and illustrate how Peierls'
bounds may be 
derived in
some long range 
systems. Unlike the very rich
ground state structures possible 
for incommensurate scalar systems (bubbles, dots, Wigner 
crystals), we prove
here that for incommensurate 
continuous spin systems
on a lattice, spiral states 
are the only ground
states. Perhaps
most noteworthy,
we perform a 
thermal stability
analysis whose results
coincide with a generalized
Mermin-Wagner inequality. 
In the general case, we
discover a new intriguing
odd-even $n$ effect
that is absent 
in conventional
ferromagnetic
or antiferromagnetic
systems: Algebraic long range order
may be possible for certain
frustrated incommensurate even
$n$ systems, while their odd $n$
counterparts exhibit an exponential decay of
correlations. Odd $n$ incommensurate
systems are generally more disordered
than their even $n$ counterparts. 
In particular, we illustrate
that many frustrated incommensurate continuous
spin systems have smectic like thermodynamics.

The outline of this article is as 
follows:
After introducing, in section(\ref{definitions}),
the terms and notations that will be used
throughout the work, we provide, in section (\ref{toy})
frustrated toy models 
which are rich enough to 
illustrate some of the 
general features that 
we aim to highlight.
These models will be employed
for illustrative purposes only.
The contents of the present publication
are not limited to these systems.
All stated in the current article
applies to all translationally
invariant systems (whether these
are the standard nearest
neighbor Heisenberg model
or the more complex frustrated
models that we introduce in 
section (\ref{toy})).
Notwithstanding their pedagogical purposes
in the current context, the toy models
presented in section(\ref{toy}) 
and myriad variants have been a subject of
much research in recent years, see e.g. 
\cite{Seul}. The physics of incommensurate orders
along with, in some instances, 
competing short- and long-range interactions
 is a rich and complex topic. 
These appear,  amongst others, in the high temperature
superconductors  \cite{Tranquada}, \cite{steve}, \cite{us}, \cite{Low},
\cite{carlson}, manganates and nickelates \cite{cheong}, 
\cite{golosov}, 
quantum Hall systems \cite{lilly}, \cite{QHE}, \cite{Fogler} 
chemical and magnetic mixtures,
crumpled membranes \cite{Seul}, and some theories of structural
glasses \cite{new}, \cite{dk}.
In many instances, there appear to be a large number
of possible ground states exhibiting nanoscale phase separation,
including stripes, labyrinths, patches, and dot arrays.
Most new results reported in this article pertain to 
incommensurate spin structures.

In section(\ref{Ising-gs}) we 
discuss the ground states of
Ising spin models and show
what patterns one should expect
in general. Once the ground states
will be touched on, we will head
on to show how Peierls bounds
may be established for many
systems having infinite range 
interactions if the 
ground states are simple. 
In section(\ref{ISING-MFT})
we shortly review mean field
solutions of the general two 
spin Ising models.
A generalization of 
the standard Peierls 
bound to other systems
with long range interactions
is provided in 
section(\ref{peierls}).
When fused with an additional
$Z_{2}$ symmetry (which
is not present in 
many of the models
that we discuss), Peierls' bounds
suffice to prove the existence
of long range order.

Henceforth, the bulk of this article focuses
on continuous spins whose number of 
components $n \ge 2$. In section(\ref{n-gs}) 
we prove that, sans 
special commensurability effects, the 
ground states of all
$O(n \ge 2)$ will typically have 
a spiral like 
structure. Section(\ref{s-stiffness}) 
details an exceedingly simple spin 
wave stiffness analysis
to gauge the effect of thermal
fluctuations on the various 
$O(n \ge 2)$ ground states. 
In section(\ref{XY-fluct}) we will discuss
thermal fluctuations within the framework
of ``soft-spin'' XY model. We will
see that the normalization constraint
gives a Dirac like equation. In the  
aftermath, the fluctuation spectrum will
be seen to match with that derived 
in section(\ref{s-stiffness}). We will
show possible links to smectic like
behavior in three dimensions.
Next, we go one step further to study the 
fully constrained ``hard-spin'' $O(2)$ and $O(3)$
models and show (in section \ref{M-W-low})
that all translationally
invariant systems in two dimensions
with an analytic rotationally
symmetric interaction kernel
never develop spontaneous magnetization.
At the end of the section our analysis
will match that of sections(\ref{s-stiffness}) and (\ref{XY-fluct})
We extend the Mermin-Wagner-Coleman bounds,
in section(\ref{M-W-I-boost}),  
to high dimensions to show
that intricacies occur if a certain
high dimensional integral will be seen to diverge. 
In the low temperature limit,
the integrand of this integral will, once again, 
match that derived by the much
more naive spin stiffness and
soft spin fluctuation analysis
of earlier sections. 
In section(\ref{n=3}) 
we will examine the ``soft-spin''
version of Heisenberg spins.
We will see that it might be
naively expected that the 
spin fluctuations in 
odd $n$ spin systems are 
larger than in those 
with an even number of
spin components. The origin of
this ``odd'' even-odd effect
is that for odd $n$ systems,
one of the spin components
is unpaired and may exhibit 
less inhibited fluctuations. 
Next, in section(\ref{n=4}), 
we carry out the spin fluctuation
analysis for four component
soft spins to see that their
spectra coincides with that
predicted in the earlier spin
stiffness analysis. We compute
the correlation functions for 
this system and find that
even $n$ incommensurate
systems may display 
algebraic long range order
in situations where their
odd $n$ counterparts of the same
system exhibit exponential decay of
correlations at all finite temperatures
(insofar as we may discern from a
perturbative analysis of soft spin models).

We match the finite $n$
analysis with its large
$n$ counterpart in
section(\ref{spherical}).
We show that in the limit
of large $n$ both odd 
and even component spin
systems behave in the same
manner. Essentially, they all
tend towards an ``odd'' behavior.
We also report on
a ``holographic'' like
effect present in 
some frustrated 
systems. In these, 
the ground state
entropy is
shown to scale 
with the surface
area of the system.
In section(\ref{high_mft}), 
we compute the critical
temperature of all 
translationally
invariant $O(n \ge 2)$ 
spin models within 
mean field theory.  
In section(\ref{brilliant}) 
we briefly remark that much of
analysis is unchanged for 
arbitrary non-translationally
invariant two spin interactions.
We conclude with a discussion
of $O(n)$ spin dynamics.

A central theme which will be repeatedly 
touched on throughout the paper
is the possibility of non-trivial
ground state manifolds. If the 
system is degenerate the effective 
topology of the low temperature
phase of the system may be classified
in momentum (or other basis). In such
instances the low temperature behavior
of the systems will be exceedingly
rich.

\section{Definitions}
\label{definitions}

We consider simple classical 
spin models of the type
\begin{eqnarray}
H = \frac{1}{2} \sum_{\vec{x},\vec{y}}
\hat{V}(\vec{x},\vec{y})[\vec{S}(\vec{x}) 
\cdot \vec{S}(\vec{y})]. 
\end{eqnarray} 
Here, the sites $\vec{x}$ and $\vec{y}$
lie on a (generally hypercubic)  
lattice of size $N$. The spins $\{ S(\vec{x}) \}$
are normalized and have $n$ components,
$\sum_{i=1}^{n} S_{i}^{2}(\vec{x}) = 1$,
at all lattice sites $\vec{x}$.
We will primarily focus on translationally 
invariant interactions $V(\vec{x},\vec{y}) 
= V(\vec{x}- \vec{y})$. 
We employ the non-symmetrical
Fourier basis convention
($f(\vec{k}) = \sum _{\vec{x}} 
F(\vec{x}) e^{-i \vec{k} \cdot \vec{x}};$
\bigskip
$ ~ F(\vec{x}) = \frac{1}{N} \sum_{\vec{k}} 
 f(\vec{k}) e^{i \vec{k} \cdot \vec{x}}$) 
wherein the Hamiltonian is diagonal and reads 
\begin{eqnarray}
H =  \frac{1}{2N} \sum_{\vec{k}} v(\vec{k}) |\vec{S}(\vec{k})|^{2}
\end{eqnarray}
where $v(\vec{k})$ and $\vec{S}(\vec{k})$ are the Fourier
transforms of $V(\vec{x})$ and $\vec{S}(\vec{x})$.
More generally, for some of the 
properties that we will illustrate,
one could consider any arbitrary real 
two spin interactions $\langle \vec{x} | V | \vec{y} \rangle$
which would be diagonalized in another basis 
$\{|\vec{u} \rangle\}$ instead of the Fourier 
basis. For simplicity, we will set the lattice constant to unity-
i.e.  on a hypercubic lattice (of side $L$) with
periodic boundary conditions the wave-vector
components $k_{l} = \frac{2 \pi r_{l}}{L}$
where $r_{l}$ is an integer (and the real space 
coordinates  $x_{l}$ are integers).
Throughout, we employ
$\vec{K}$ to denote reciprocal 
lattice vectors and
$\Delta (\vec{k})$  as a shorthand 
for the lattice lattice Laplacian:
\begin{eqnarray} 
\Delta(\vec{k}) = 2 \sum_{l=1}^{d} (1-\cos k_{l}). 
\end{eqnarray}
In some of the frustrated systems that we will soon consider,  
$v(\vec{k})$ may be written explicitly as 
the sum of several terms: those favoring 
homogeneous states $\vec{k} \rightarrow 0$, 
and those favoring zero wavelength 
$\vec{k} \rightarrow \infty$ 
(or $\vec{k} \rightarrow (\pi,\pi,...,\pi)$
on a lattice.) As a result of this competition,
modulated structures arise on an intermediate scale.

\section{Toy Models}
\label{toy}

Although we will keep the discussion 
very general, it might be useful to have 
a few explicit applications in mind.
There is a lot of physical intuition 
which underlies the upcoming models. Unfortunately,
here they will
merely serve as nontrivial toy models 
on which we will able to exercise 
our newly gained intuition.
The systems to be presented
are frustrated: not all two
spin interactions can be 
simultaneously satisfied. 
They entail competing interactions.
Such systems appear, amongst others, in the high temperature
superconductors, CMR materials,
quantum Hall bars, chemical and magnetic mixtures,
crumpled membranes, and some theories of structural
glasses \cite{Seul}. Toy models in which these issues can be 
addressed are useful. In the current context, we choose these 
examples as they highlight subtleties typically 
absent in the more standard spin models.
In the continuum limit, any incommensurate rotationally 
symmetric model (with an interaction $v(\vec{k})$ which is 
analytic about its minima and having non-vanishing second derivatives) 
will share much the same physics as the two specific models 
discussed below. When any incommensurate
spin system possesses a rotational symmetry,
the minimizing manifold of $v(\vec{k})$ 
is a $(d-1)$ dimensional shell of
radius $q>0$.

{\bf{The Coulomb Frustrated Ferromagnet}}

We now introduce the ``Coulomb Frustrated Ferromagnet''.
This is a toy model of a doped Mott insulator \cite{steve}, \cite{Low}, 
of phase separation in High Landau level Quantum Hall 
systems \cite{QHE}, \cite{Fogler} and of
certain amphiphilic systems. It has been argued that
within the Mott insulator the tendency, of holes, to phase
separate at low doping is  
frustrated, in part, by electrostatic repulsion \cite{steve}.
In three dimensions, a simple spin Hamiltonian \cite{us}
which represents these 
competing interactions is 
\begin{eqnarray}
H_{Mott} = - \sum_{\langle \vec{x},\vec{y} \rangle} S(\vec{x}) S(\vec{y}) +  
\frac{Q}{8 \pi} \sum_{\vec{x} \neq \vec{y}} \frac{S(\vec{x}) 
S(\vec{y})}{|\vec{x}-\vec{y}|} \nonumber
\\ = \frac{1}{2N} 
\sum_{\vec{k}} [\Delta(\vec{k})+ \sum_{K} |\vec{k}-\vec{K}|^{-2}]
 |S(\vec{k})|^{2}.
\end{eqnarray}
In the second line we employed the Poisson summation
formula. Here $\{\vec{K}\}$ is the set of all reciprocal
lattice vectors as introduced in Section(\ref{definitions}).
Here, $S(\vec{x})$ is  a coarse grained scalar variable which represents the 
local density of mobile holes. Each site $\vec{x}$ 
represents a small region of space in which
$S(\vec{x})>0$, and $S(\vec{x})<0$ correspond to hole-rich and 
hole-poor
phases respectively.  
In this Hamiltonian, the first ``ferromagnetic''
term represents the short-range
(nearest-neighbor) tendency of the holes to phase-separate and form 
a hole-rich ``metallic'' phase, whereas the frustrating effect of the 
electrostatic repulsion between holes is present in the second term.
Non-linear terms in the full Hamiltonian typically fix the locally
preferred values of $S(\vec{x})$. One may consider
$d \neq 3 $ dimensional variants wherein the spins 
lie on a hypercubic lattice, and the Coulomb kernel
in $H_{0}$ is replaced by $\frac{Q}{2 \Omega_{d}}|\vec{x}-\vec{y}|^{2-d}$ 
(or by $[\frac{Q}{4 \pi}~ \ln |\vec{x}- \vec{y}|~]$ in two dimensions 
\cite{2dCoulomb})
where $\Omega_{d} = 2 \pi^{d/2}/\Gamma (d/2)$.
Here the competition between both terms, when $Q \ll 1$ 
favors states with wave-numbers $\simeq  Q^{1/4}$.
The introduction of the Coulomb interaction is manifestly
non-perturbative: it is long range. Moreover,
the previous ferromagnetic ground state becomes, 
tout a' coup, infinite in energy. 
We will, for the most part, focus on the 
continuum limit of this Hamiltonian 
where the kernel
becomes
\begin{eqnarray}
v_{Mott}^{cont}(\vec{k}) = Q k^{-2}+ k^{2}(1+\sum_{\vec{K} \neq 0} 
\frac{3}{K^{4}})
\end{eqnarray} 
After rescaling,
this may also be regarded as the small $\vec{k}$ 
limit of the more general 
\begin{eqnarray}
v_{Q}(\vec{k})= \Delta(\vec{k}) + Q [\Delta(\vec{k})]^{-1} +
A [\Delta(\vec{k})]^{2} \nonumber
\\ + \lambda \sum_{i \neq j} (1- \cos k_{i})(1-\cos k_{j}) + O(k^{6}).
\label{Q}
\end{eqnarray}
The constants $A$ and $\lambda$ are pinned
down if we identify $v_{Q}(\vec{k})=v_{Mott}(\vec{k})$.
Here, we will modify them
in order to streamline the quintessential
physics of this system. First, we set $A=0$: In the continuum
limit this term is not large nor does it
lift the ``cubic rotational
symmetry'' \cite{explain_cubic} present to lower 
order. Next, we allow $\lambda $ to vary in order to 
turn on and off ``cubic rotational symmetry''
breaking effects \cite{explain_cubic}.
The kernel $v_{Q}(\vec{k})$ may be regarded as 
$v_{Mott}$ augmented by all possible next to nearest
neighbor interactions. As our Hamiltonian respects
the hypercubic point symmetry group,
by surveying all possible
values of $\lambda$ 
we should be able to make
general statements regarding the
possible phases
(within the planes) 
of real doped Mott insulators. In dimension $d=2$, whenever 
$\lambda >0$, the minimizing wave-vectors will lie along the 
cubic axis and ``horizontal''  order
will be expected. When $\lambda <0$ 
the minimizing wave-vectors lie along
the principal diagonals and diagonal order 
is expected. At large values of $Q$,
when the continuum limit no longer
applies, trivial extensions of
these minimizing modes are
encountered where one or more of
the wave vector components is
set to $\pi$. The work reported
in this paper focuses 
on spin systems on 
a lattice. The situation
becomes far richer 
in continuum field theories.
In high level Quantum Hall systems,
calculations on a similar
model indicate \cite{Fogler} 
that the system undergoes a sequence 
of transitions. At very low 
excess filling fractions, $\nu - N< 1/N$
(with $N$ the integer part of the filling $\nu$),
a Wigner crystal of the cyclotron orbit centers
is formed. As $\nu$ is further increased,
transitions between bubbles of increasing
size occur. Ultimately, at a sufficiently large filling 
$\nu$, stripes appear. Similar textures are observed 
in chemical mixtures \cite{Seul} which may be
described by similar models \cite{Kawasaki}.
We will prove, however, that this series 
of transitions between various textures does 
not occur for multi-component spins on a lattice.
On a lattice, barring special commensurate
points that we detail below (e.g. 
the standard uniform and Neel phases and 
other, more intricate, commensurate states), 
only spiral (stripe-like)
states may appear. 

If the ferromagnetic system were 
frustrated by a general long range kernel
of the form  $V(|\vec{x}-\vec{y}|) \sim |\vec{x}-\vec{y}|^{-p}$
we could replace the $[\Delta(\vec{k})]^{-1}]$ in $v_{Q}(\vec{k})$
by the more general $[\Delta(\vec{k})]^{(p-d)/2}$.
For instance, in a spin-lattice version
of the Quantum Hall problem, $p=1$ and
$d=2$. Here,  in the continuum limit, the minimizing modes
 are $q \sim Q^{1/(2+d-p)}$ and as 
the reader will later be able easily verify all 
our upcoming analysis can be reproduced
for any generic long range frustrating interaction 
with identical conclusions.

\begin{figure}
\centering
\hspace{0.0in}{\psfig{figure=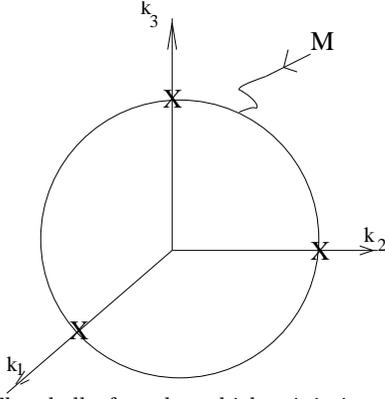,width=2.0in,clip=}}
\caption{The shell of modes which minimize the 
energy in the continuum limit. For the case just discussed
This sphere is of radius $|\vec{q}|=Q^{1/4}$.}
\label{fig:Minimizing_manifold_1}
\end{figure}

In Fig.(\ref{fig:Minimizing_manifold_1}), we schematically depict
the manifold ($M$) of the minimizing modes
in $\vec{k}$ space. When no symmetry
breaking terms ($ \lambda =0$) are present,
in the continuum
limit it is $M$ is the surface of sphere of radius $Q^{1/4}$.
If $\lambda \neq 0$ 
this degeneracy will be lifted:
only a finite number of modes will minimize the energy.
When $\lambda >0$ there will be $2d$ minimizing modes (denoted by the big X
in the figure) along the coordinate axes. In the 
up and coming we will focus mainly
on $ \lambda \ge 0$.
When $\lambda <0$, a moment's
reflection reveals that there will be $2^{d}$
minimizing modes along the diagonals,
i.e. parallel to $(\pm 1,\pm 1,\pm 1)$
(and in this case, they will have 
a modulus which differs 
from $Q^{1/4}$). Unless explicitly stated otherwise, 
we will set $\lambda=0$
for calculational convenience
and when a finite $\lambda$ is
invoked it will be made 
positive (to avoid the $\lambda$ 
dependence of $|\vec{q}|$ incurred
when the former is negative).
At times, we will present results
for $v_{Q}(\vec{k})$ at sizable $Q$,
even though the model was motivated as
a good caricature of $v_{Mott}$
only in the continuum limit (at 
small wave-vectors  $q \sim Q^{1/4}$).

{\bf{Membranes}}

In several
fluctuating membrane systems, 
the affinity of the molecular 
constituents (say A and B) 
for regions of different
local curvature frustrates
phase separation \cite{Seul}.
Let us define $S(\vec{x})$ to be the difference between 
the A and B densities at $\vec{x}$. 
In the continuum, the energy of the system contains a contribution,
\begin{eqnarray}
H_{mix} = \frac{b}{2} \int d^{2}x~|\nabla S|^{2} 
\end{eqnarray}
reflecting the demixing of A and B species.
Instead of considering long-range interactions,
we now allow for out-of-plane (bending) distortions of the sheet. 
Specifically, we assume that the two molecular 
constituents display an affinity
for regions of different local curvature of the sheet. 
This tendency can be modeled by introducing a 
coupling term between the local composition $S(\vec{x})$
and the curvature of the sheet.
If the distortions remain small,
we may write 
\begin{eqnarray}
H_{c} = \int d^{2}x~ [\frac{1}{2} \sigma
 |\nabla h(\vec{x})|^{2}+\frac{\kappa}{2}[\nabla^{2} h(\vec{x})]^{2}+
\Lambda S(
\vec{x}) \nabla^{2} h(\vec{x})]
\nonumber
\\
\equiv \int d^{2}x {\cal{H}}_{c},
\nonumber
\end{eqnarray}
where $h(\vec{x})$ represents the height profile of the sheet
(relative to a flat reference state),
$\sigma$ is its surface tension , and $\kappa$ 
is its bending modulus; $\Lambda$, the coefficient of the 
last term in the expression measures the strength of the coupling 
of the local curvature $\nabla^{2}h$ and the local composition 
$\phi$, which we have included here to lowest (bilinear) order.
This coupling term reflects the different affinities of the molecular 
constituents A $(S =1$ corresponds to pure A composition)
and B ($S =0$ corresponds to pure B composition) for, respectively,
 convex ($\nabla^{2}h>0$)
and concave ($\nabla^{2}h<0$) regions of the interface. 
The variational Eqs. for the total energy $H=H_{\phi}+H_{c}$,
with respect to the membrane shape $\{h(\vec{x})\}$
read
\begin{eqnarray}
-\sigma \nabla^{2} h + \kappa \nabla^{2} (\nabla^{2} h)+
 \Lambda \nabla^{2} S =0.
\end{eqnarray}
If $|\kappa \nabla^{2}(\nabla^{2} h)| \ll
 \min \{ |\sigma \nabla^{2} h|,|\Lambda \nabla^{2} S| \}$ 
then 
\begin{eqnarray}
\Lambda S(\vec{x}) = \sigma h(\vec{x}) + g(\vec{x}),
\end{eqnarray}
with $g(\vec{x})$ a harmonic function
satisfying boundary conditions.
In ${\cal{H}}_{c}$, after an 
integration by parts,
\begin{eqnarray}
\Lambda S \nabla^{2} h \approx \Lambda S
 \frac{\Lambda}{\sigma} \nabla^{2} S
\rightarrow -\frac{\Lambda^{2}}{\sigma}(\nabla S)^{2}.
\end{eqnarray}
\begin{eqnarray}
H_{mix}+H_{c} \approx \int d^{2}x~[\frac{1}{2} b^{\prime} |\nabla S|^{2} +
 \frac{\Lambda^{2} \kappa}{2 \sigma^{2}} (\nabla^{2}S)^{2}],
\end{eqnarray}
where $b^{\prime} \equiv b-\frac{\Lambda^{2}}{\sigma}$.
This effective Hamiltonian, which is a function of $S$ alone, reads
\begin{eqnarray}
H = \int d^{2}k~ v_{membrane}(\vec{k}) |S(\vec{k})|^{2}, 
\end{eqnarray}
where $v_{membrane}(\vec{k}) = 
\frac{b^{\prime}}{2}k^{2}+\frac{\Lambda^{2} \kappa}
{2 \sigma^{2}}k^{4}$ is the 2D Fourier transform. 
A negative $b^{\prime}$ obtained when $b<\Lambda^{2}/\sigma$,
signals the onset of a curvature instability of the sheet.
This instability generates a pattern of domains that differ in
composition as well as in local curvature and thus assume 
convex or concave shapes.
The characteristic domain size corresponds to the existence of 
the minimum of the free energy at a non-zero wave number.
The modulation length $d \sim q^{-1} \simeq
\sqrt{(\Lambda^{2} \kappa/\sigma^{2})/|b^{\prime}|}$
with $q=|\vec{q}|$ the minimizing wavenumber modulus 
of $v_{membrane}(\vec{k})$.
After scaling, this model may be regarded as the continuum version of the 
frustrated short
range kernel
\begin{equation}
  v_{z}(\vec{k}) = z \Delta^{2}(\vec{k}) - \Delta (\vec{k}),
\end{equation}
(where $z= -\Lambda^{2}\kappa /(\sigma^{2} b^{\prime})$)
on the lattice.
The real-space lattice Laplacian
\begin{equation}
  \langle \vec{x}| \Delta |\vec{y} \rangle  = \left\{ \begin{array}{ll}
      2d & \mbox{ for $\vec{x}=\vec{y}$} \\
      -1 & \mbox{ for $|\vec{x}-\vec{y}| = 1$}.
\end{array}
\right.
\end{equation}
All matrix elements $\langle \vec{x}| \Delta^{R} |\vec{y} \rangle = 0$ 
for $|\vec{x}-\vec{y} |> R$ = Range.
Our system is of $Range=2$.
Explicitly,
\begin{eqnarray}
  \langle \vec{x}| \Delta^{2}| \vec{y} \rangle \mbox{ } = \mbox{ }&& 2d(2d+2) \mbox{
    for } \vec{x} = \vec{y} \nonumber
  \\
  \ \ && -4d \mbox{ for  } |\vec{x}-\vec{y}| = 1 \nonumber \\
  \ \ && 2 \mbox{ for } (\vec{x} -\vec{y}) = 
(\pm \hat{e}_{\ell} \pm \hat{e}_{\ell^{\prime}}) \mbox{ where  }
 \ell \neq \ell^{\prime}      \nonumber \\
  \ \ && 1 \mbox{ for a $ \pm 2 \hat{e}_{\ell} $ separation}.
\end{eqnarray}
We will extend the investigation of
this model over a broader range
of parameters than suggested 
by its initial physical motivation.
In the continuum limit, theories 
with high order derivative terms
generally give rise to 
$v(\vec{k})  = P(k^{2})$.  
where $P$ is some polynomial.
Although $v_{z}(\vec{k})$ and
its likes are  artificial
on the lattice, their continuum 
limit is quite generic.
Later on  we will show that if 
$P(k^{2})$ attains its global
minima at finite $|\vec{k}|$,
then thermal instabilities
can incur extremely 
low values of $T_{c}$.

\section{Ising Ground States}
\label{Ising-gs}

In most sections to follow, the bulk of 
our reported results pertain to continuous
spins. We now, however, present a short overview
of discrete spin systems in order present new results 
and to illustrate how Ising systems may be addressed within a 
similar momentum space framework. Towards the end of this section, 
will establish the existence of a new ``holographic'' ground state
entropy in certain incommensurate Ising systems wherein the ground
state entropy scales as the surface area of the system.

In an ``Ising''  system $S(\vec{x})= \pm 1$
at all lattice sites $\vec{x}$. Stated alternatively, 
the scalar ($n=1$) spins satisfy a normalization
constraints $\{ S^{2}(\vec{x}) =1 \}$
at all $N$ lattice
sites $\vec{x}$. Henceforth, we will adopt
the latter point of view. 
The set of minimizing wave-vectors
 $\vec{q}$, $
v(\vec{q} \in M) \equiv \min_{\vec{k}} \{ v(\vec{k}) \}$
defines a manifold $M$.
If the local normalization constraints are swept
aside then it is clear that the ground states
are superpositions of sinusoidal waves with
wave-vectors $\vec{q} \in M$. One would 
expect this to be true, in spirit, also in
the highly constrained Ising case,
if $v(\vec{k})$ is sharply dipped at 
its global minima.  If, in a non-rigorous 
setting, we ``digitize'' 
a particular plane wave 
$S(\vec{x}) = sign(\cos(\vec{q}_{1} \cdot \vec{x}))$
and compare it with the exact (numerical) 
ground state, then we will find encouraging agreement
in certain cases. For instance, this gives reasonable
accord when $H= H^{Mott}$ \cite{thesis,us,Low}. 
This Hamiltonian (with some twists) was investigated
in \cite{steve} on a square ($d=2$) lattice.
In the continuum limit 
(i.e. if the lattice is thrown
away) we might naively anticipate a huge
ground state degeneracy- 
a ``digitized plane wave'' 
for each wave vector $\vec{q}$ lying on
the $(d-1)$ dimensional manifold
 $\{ M_{Q} : q^{4} = Q \}$
This large degeneracy 
might give rise to
a loss of stability 
against thermal fluctuations.
Striped phases
(i.e. ``digitized plane waves'') were found
in virtually all
of the parameter range \cite{thesis,us,Low}.
Only for a very small range of parameters
were more complicated periodic structures
found.

An intuitive feeling can be gained by considering 
a one dimensional pattern such as 
\begin{eqnarray}
++--++--++--...
\end{eqnarray}
This pattern is
a pure mode $S_{period=4}(x) = \sqrt{2} \cos[\frac{\pi}{2} x - 
\frac{\pi}{4}]$. With this observation at hand, a double 
checkerboard pattern such as 
\begin{eqnarray}
 ++--++-- \nonumber
\\ ++--++-- \nonumber
\\ --++--++ \nonumber
\\ --++--++ 
\end{eqnarray}
extending in all directions in the plane is given by
\begin{eqnarray}
S(\vec{x}) = 2 \cos[\frac{\pi}{2} x_{1} - 
\frac{\pi}{4}] \cos[\frac{\pi}{2} x_{2} - \frac{\pi}{4}] = \nonumber
\\  \cos[\frac{\pi}{2}(x_{1}+x_{2})-\frac{\pi}{2}] +
 \cos[\frac{\pi}{2}(x_{1}-x_{2})].
 \end{eqnarray}
Such a $ 4 \times 4 \times 4$ periodic pattern in 
three dimensions would include the eight modes 
$\frac{1}{2} (\pm \pi,\pm \pi . \pm \pi)$. 
This example illustrates an simple premise. 
A system composed of a periodic building block
whose dimensions $p_{1} \times p_{2} \times p_{3}$, is given by
\begin{eqnarray}
S(\vec{x}) = S_{p_{1}}(x_{1}) S_{p_{2}}(x_{2}) S_{p_{3}}(x_{3}).
\label{ppp}  
\end{eqnarray}
If a configuration $S_{p}(x)$ contains the modes $\{ k_{p}^{m} \}$
with amplitudes $\{ S_{p}(k_{m}) \}$, then Fourier transforming 
the periodic configuration $S(\vec{x})$ one will find  the modes
$(\pm k_{p_{1}}^{m_{1}}, \pm k_{p_{2}}^{m_{2}}, \pm k_{p_{3}}^{m_{3}})$
appearing with a weight $ \sim |S_{p_{1}}(k_{1}) \times S_{p_{2}}(k_{2})
\times S_{p_{3}}(k_{3})|^{2}$. For high values of the periods $p$, 
the weight gets scattered over a large set of wave-vectors. 
If $v(\vec{k})$ has sharp minima, such states will not be favored.
The system will prefer to generate patterns s.t. in all 
directions $i$ albeit one $p_{i}=1 ~~(\mbox{or perhaps~ }2)$. For a 
$p_{1} \times p_{2} \times p_{3}$ repetitive pattern, the discrete
Fourier Transform will be nonzero for only $\prod_{i=1}^{3} ~p_{i}$
values of $\vec{k}$. This trivial observation  suggests
the phase diagram obtained by \cite{Low} in two 
dimensions. The intuition is obvious.
We have derived \cite{us},  rigorously, the ground states
in only several regions of its parameter space
(those corresponding to ordering with half a reciprocal
lattice vector),  and on  a few 
special surfaces (corresponding to ordering 
with a quarter of a reciprocal lattice vector).
In all of these cases the Ising states may
be expressed as superpositions of the
lowest energy modes
$\exp[i \vec{q}_{m} \cdot \vec{x}]$.
Lately, a beautiful extension was carried out
by \cite{Gilles}. Another nice work,
focusing on scalar fields in the
continuum is \cite{muratov}.
In the current publication,
however, we are concerned only
with spins on
a lattice.  We now ask whether commensurate lock-in is to be expected.
The energy of the Ising ``digitized plane wave''
on an $L \times L \times L$ lattice where
$\vec{q} = (q_{1},0,0)$ with $q_{1} = 2 \pi/m$, with even $m$, reads
\begin{eqnarray}
E = \frac{1}{2N} \sum_{\vec{k}} v(\vec{k}) |S(\vec{k})|^{2} =
\nonumber
\\ \frac{8}{m} \sum_{j=1,3,...,m-1}
 \frac{v(\vec{k}=(\frac{2 \pi j}{m},0,0))}{|\exp[2 \pi i j/m]-1|^{2}} =
\nonumber
\\  \frac{2}{m} 
\sum_{j=1,3,...,m-1}
\frac{v(\vec{k}=( \frac{2 \pi  j}{m},0,0))}{\sin^{2}(\pi j/m)}. 
\end{eqnarray}
The lowest energy state amongst all
states of the form considered 
is a possible candidate
for the ground state.

Unlike the previous 
paragraphs and those
to follow, we now try to present
the reader with
an intuitive feeling.
The following paragraph 
is strictly non-rigorous.
For the particular model long-range introduced 
in Eq.(\ref{Q}), 
it seems that for small values of $Q$, it might be 
worthwhile to have an incommensurate phase. This is,
in a sense, obvious- all low energy modes are of very small
wave-number $q$ and hence not of low commensurability. 
In a stripe phase having of size $2 \pi|\vec{q}_{ground}|^{-1}$,
the energy
\begin{eqnarray}
E = \frac{1}{N} \sum_{n=0}^{\infty}
 \frac{16v(\vec{k}=(2n+1)\vec{q}_{ground})}
{(2n+1)^{2}
\pi^{2}}.
\end{eqnarray} 
In the example of Eq.(\ref{Q}) for $q_{ground} \ll 1$, 
the higher harmonics
$\vec{k} = (2n+1) \vec{q}$, do not entail high energies. 
For large values of $Q$, $q_{ground}  \simeq O(1)$,
and $v(\vec{k}=(2n+1)\vec{q}_{ground})$ can be very large if $[(2n+1)\vec{q}_{ground}]$
approaches a reciprocal lattice vector $\vec{K}$.
Under these circumstances
it will pay off to have a commensurate structure; for a 
$u_{1} \times u_{2} \times ... \times u_{d}$ repetitive block only the modes 
$\vec{k} = 2 \pi(\frac{n_{1}}{u_{1}},\frac{n_{2}}{u_{2}},...,
\frac{n_{d}}{u_{d}})$
will be populated (i.e. have a non-vanishing
$|\vec{S}(\vec{k})|^{2}$)- the ferromagnetic point (a reciprocal 
lattice point)
will not be approached arbitrarily close- if that is not true
weight will be smeared over energetic modes. Generically, we will
not be expect commensurate lock-in in $\lim \vec{q} \rightarrow 0$
for {\it{any}} theory with a  frustrating long range
interaction. Although we have considered only 
striped phases (which have previously argued are the only
ones generically expected), it is clear that this 
argument may be reproduced for more exotic 
configurations (such as bubbles, dots, cylinders
etc.) As an aside, by replacing 
Fourier lattice sums by integrals, we note 
that in the continuum limit of $H_{Mott}$, 
the ground state wave-vector in three dimensions, 
is immediately seen to be 
$q_{ground} \sim Q^{1/3}$ \cite{one_third}- 
not the value suggested by dimensional
analysis (if the lattice constant units
are ignored) -
the minimizing wavenumber $q \sim Q^{1/4}$. 
We will prove that for continuous spins $q_{ground} =q$.
Similar results are obtained for the full Hamiltonian $H_{Mott}$
(not only its continuum limit) and related systems, 
e.g. \cite{Gilles}, \cite{Kawasaki}, \cite{choksi}.

For finite range interactions it is easy to
prove, by covering the system with large maximally
overlapping blocks, that there will be a 
regime about $\vec{q} = 0$, for which we
will find the ferromagnetic 
ground state. A polynomial in $\Delta(\vec{k})$ 
will have its minima at $\Delta(\vec{k}) = const$, i.e.
on a (d-1) dimensional hypersurface(s) in $\vec{k}$-space or
at the (anti)ferromagnetic point.
The kernel $v_{z}(\vec{k})= z \Delta^{2} - \Delta$
has its minima ($z > 0$) at $ \vec{q} \in M_{z} : 
\Delta (\vec{q}) = \min \{ \frac{1}{2z} \mbox{,    } 4d \}$.
For $ z > \frac{1}{8d}$: $M_{z}$ is $(d-1)$ dimensional.
We may divide the lattice into all maximally over-lapping 
$5 \times 5 \times ...\times 5$
hyper-cubes centered about each site of the lattice.
\begin{equation}
  Energy = \frac{1}{5 \times 6^{d-1}} \sum_{hypercubes} \epsilon(hypercube)
\end{equation}
and evaluate the energies $\epsilon$ of all $5 \times 5 \times... \times 5$
Ising configurations. Of all $2^{5^{d}}$
configurations the Neel state will have the lowest energy for a
sliver about $z=\frac{1}{8d}$. Analogously for $z > z_{top}>>1$,
by explicit evaluation, the ground state will
be ferromagnetic. Contour arguments can be employed
and a finite lower bound on $T_{c}$ generated.

If such a system 
satisfies periodic boundary 
conditions along the two square
lattice diagonals 
$ \hat{e}_{\pm} \mbox{ :}$ defined by $ x_{1} \mp x_{2} = const$,
(with $x_{i}$ the Cartesian coordinates), then
an exponentially large number of Ising ground states can be 
constructed whenever $z = \frac{1}{8}$.
To illustrate this, we note that when $z=\frac{1}{8}$,
all minimizing modes lie on
\begin{equation}
  \vec{q} \in M_{z=\frac{1}{8}} : |q_{1} \pm q_{2}| = \pi.
\end{equation} 
By prescribing an arbitrary spin configuration along $x_{-}$
\bigskip
 and fixing $S(x_{-},x_{+}) = S(x_{-},0) (-1)^{x_{+}}$:
\begin{equation}
  {S(\vec{k}) = \sum_{x_{-}} S(x_{-},0) \exp(ik_{-}x_{-}) \sum_{x_{+}}
    (-1)^{ x_{+}}\exp(ik_{+}x_{+})}
\end{equation}
vanishes for $|k_{+}| = |k_{1}+k_{2}| \neq \pi$.  Similarly, by taking
the transpose of these configurations we can generate patterns having
$S(\vec{k}) = 0$ unless $|k_{-}| = |k_{1}-k_{2}| = \pi$.  The ground
state degeneracy is bounded from below by, the number of independent
spin configurations that can be fashioned along $x_{+} \mbox{ or }
x_{-} , (2^{L+1}-2)$ where $L$ is the length of the system along the
$x_{\pm}$ axis. The number of $\vec{q}$ values, commensurate with the
diagonal periodic boundary conditions, lying on $M_{z=\frac{1}{8}}$ is
$(4L-2)$. We have thus proved that the ground state
entropy in this two dimensional
system is, at least, linear in the perimeter of the
system. In a later section, we will later demonstrate 
that in the large $n$ limit, such ``holographic'' like
entropies do generically arise whenever
the minimizing manifold $M$ is (d-1) dimensional. 
The Ising ground state degeneracies are bounded
from above by those of the spherical
model (as any Ising configuration
is a viable configuration in the
spherical model). As the ground
state entropy for our model
is bounded both from above
and from below by the perimeter
of the system, the ground state
entropy rigorously scales 
as the system size $L$.

\begin{figure}
\centering
\hspace{0.0in}{\psfig{figure=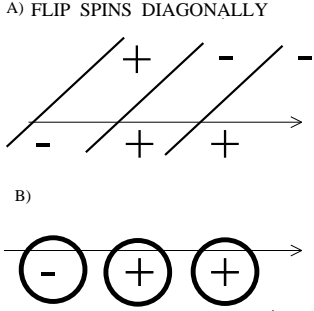,width=1.7in,clip=}}
\caption{Simple ground states for $v_{z}(\vec{k})$ with $z=1/8$
(and for $v_{Q}(\vec{k})$ when $Q=16$): A) Their construction- 
along the arrowed line we arbitrarily
prescribe spins. For each move along the dotted diagonal lines
we flip the spins. B) In this chain of
diagonal super-spins there
is  no stiffness against 
flipping.
}
\label{fig:diagonal_ground_states}
\end{figure}

Similarly, in dimensions $d>2$, we can set $(d-2)$ of the $\vec{q}$ components
to zero. There are $d(d-1)/2$ cross-sections of the d-dimensional
$M_{z=\frac{1}{8}}$, all looking like the the two-dimensional $M$ just
discussed (i.e.  $|q_{1} \pm q_{2}| = \pi$ ).  The real-space ground
state degeneracy is bounded from below by $d(d-1)[2^{L}-1]$ (along the
$(d-2)$ zero-mode directions the ground state spin configurations
display no flip).
If we regard each diagonal row of
spins as a ``super-spin'' then  we will
see that flipping any ``super-spin'' entails
no energy cost. This is reminiscent to 
a nearest neighbor Ising chain where
the energy cost for flipping a spin
is dwarfed by comparison to the
(logarithmically) extensive 
entropy. We might expect that here, too,
ordering might be somewhat inhibited.
In two dimensions, 
\begin{equation}
  \lim _{z \rightarrow \infty} M_{z} : {\vec{q}}^{\mbox{ } 2} =
  \frac{1}{2z},
\end{equation}
the ``average'' number of allowed $\vec{q} \in M_{z} \mbox{ values }
\ll O(L)$ (and similarly for the onset $\lim_{z \rightarrow
  \frac{1}{8d}^{+}} M_{z} : (\vec{q}-(\pm \pi,\pm \pi))^{2} =
(\frac{1}{8d}-\frac{1}{2z})$).  For a hypercubic lattice of
size $L_{1} \times L_{2} \times ... \times L_{d}$
in  $d>2$  many discrete reciprocal points will give rise to the 
same value of $\Delta(\vec{k}) \sim \vec{k}^{2}$. The proof is trivial:
if all $L_{l} = L$, then the number of possible $\vec{k}^{2}$
values is bounded by $dL^{2}$, whereas there are $L^{d}$~ $\vec{k}-$values.
Therefore, on ``average'', the number of $\vec{k}$ points lying on $M_{z}$,
or more precisely lying  lying the closest to $M_{z}$, s.t. $|\Delta(\vec{k}) -
 \frac{1}{2z}|$
is min, is, at least, $O(L^{d-2})$. 
Of these, $\frac{2^{d-z}~d!}{\Pi_{i}(n_{i}!)}$ wave-vectors,
 with $n_{i}$ (and $z$)
denoting the number of identical components (and the
number of zero components)
of a certain $\vec{k}_{1}$ nearest to $M_{z}$,
are related to $\vec{k}_{1}$ by symmetry.
As we have stated previously, in the continuum limit ($q \rightarrow 0$)
any short range kernel (including this one) will have a uniform 
(ferromagnetic) ground state.
 However the impossibility
of constructing ground states that contain 
only ``good'' Fourier modes $S( \vec{k} \in M)$ when
$M$ shrinks to a curved surface enclosing the origin
is more general and will proved in the next section. 
For this short range model, even
for $ z \neq \frac{1}{8}$
a  huge ground state degeneracy is expected. 
A ``plane wave''  might correspond to each
wave-vector $\vec{q}$ (or commensurate wave-vectors 
nearby) lying on the $(d-1)$ dimensional manifold $M$.
As we shall prove later on, even in high dimensions, 
and even if the interactions are long ranged,
in the continuum limit it will not be possible 
to construct Ising states
in which $S (\vec{k} \not\in M) =0$ unless 
the minimizing manifold $M$ contains 
flat non-curved segments (or more 
generally intersects a plane
at many points).

\section{A universal Peierls bound}
\label{peierls}

In this section, we illustrate how
a Peierls' bound can be obtained
in a host of systems with
two spin interactions. Although
we specialize to translationally
invariant two spin interaction
in this section, as we will
show argue later, 
the technique that we outline
below can be extended to 
non-translationally invariant
systems as well. In Peierls' elegant proof 
of long range order in the two and 
higher dimensional Ising
models at sufficiently
low temperatures, one of the 
necessary ingredients
is the Peierls' bound.
This bound amounts to 
the demonstration that
having one ground state 
domain surrounded by another
costs an energy which 
is, at least, linear
in the length of
their interface.
In this section, we prove 
that such a bound
is quite generic
to many long range
systems. Albeit suggestive,
the bound, on its own, does not
suffice to demonstrate
long range order.
The unfamiliar reader
may peruse \cite{PEIERLS}.
If a real hermitian kernel $v(\vec{k})$ attains
its minima in only
a finite number of commensurate
reciprocal lattice points
$\{\vec{q}_{i}\}$,
then a Peierls bound can, in some instances,
be proven for an infinite range model:
When possible this is suggestive of a
finite $T_{c}$.  
For instance, the bound for a (lattice) Coulomb gas 
(with the kernel solving the
discrete Laplace equation on the lattice) is trivially
generated. 
\begin{eqnarray}
v_{e}(\vec{k}) - v_{e}(\vec{q}) = e/\Delta(\vec{k}) -
e/\Delta(\vec{q} = \pi,\pi,\pi) \nonumber
\\ \ge -A (\Delta(\vec{k}) -
\Delta(\vec{q} = \pi,\pi,\pi)) ),
\label{ve}
\end{eqnarray}
with $A = 16 d^{2}$. Here, the Hamiltonian 
\begin{eqnarray}
H = \frac{1}{2N} \sum_{\vec{k}} 
v_{e}(\vec{k}) |S(\vec{k})|^{2} \equiv 
\frac{1}{2N} \sum_{\vec{k}} \frac{e}{\Delta(\vec{k})} |S(\vec{k})|^{2}
\end{eqnarray}
depicts a lattice Coulomb interaction
(wherein the spins portray charges).
The right hand side of Eq.(\ref{ve}) is the
kernel of an antiferromagnet. Both system share the same ground
states. For a given configuration the energy penalty for the Coulomb
gas 
\begin{eqnarray}
\Delta E_{e}= 1/(2N) \sum_{\vec{k}} [v_{e}(\vec{k})-v_{e}(\vec{q})]
|S(\vec{k})|^{2}   
\end{eqnarray}
 is bounded from below by the corresponding penalty
in an antiferromagnet of strength $A$.  In $d=2$ the contour penalty
of the antiferromagnet is $ 2A|\Gamma|$, ($|\Gamma| \equiv$ length of
the contour $\Gamma$). A similar trick may frequently 
be employed when the minimizing wave-vectors attain other 
commensurate values. It relies on comparison to a short
range kernel for which a Peierls bound is trivial. 
This can be extended quite generally.
All translationally invariant system
with commensurate minimizing 
wave-vectors 
($\vec{q} = (0,0, \ldots 0), (\pi, 0, \ldots 0), \ldots$)
at which the minimum of $v(\vec{k})$ at $\vec{k}=\vec{q}$
is quadratic may be bounded by a kernel 
of a system having nearest neighbor ferromagnetic 
and antiferromagnetic bonds for which the Peierls bound is trivial
(linear in the perimeter of the domain
wall $\Gamma$).

\section{Ising Weiss Mean-Field Theory}
\label{ISING-MFT}
We now estimate the critical temperature
for all incommensurate spin system in which
within mean-field. If, in these systems, 
when $T<T_{c}$, the on-site magnetization
$ \langle S(\vec{x}) \rangle = 
s~ sign[\cos(\vec{q} \cdot \vec{x})]$ (as 
suggested by some of the examples
discussed hitherto), then the sum
\begin{eqnarray}
\sum_{\vec{y}} \langle S(\vec{y}) \rangle V(\vec{x}=0,\vec{y}) =
\frac{1}{N} \sum_{\vec{k}} \langle S \rangle (\vec{k}) v(-\vec{k}) \nonumber \\
= s \sum_{n=0}^{\infty} \frac{4(-1)^{n}}{(2n+1)\pi}[v(\vec{k} 
= (2n+1) \vec{q})) 
+ v(\vec{k} = -(2n+1) \vec{q})],
\nonumber
\end{eqnarray}
where we assumed $q_{i} = \frac{t_{i}}{u_{i}}$ with $u_{i} \gg 1$ for all 
$d$ components in replacing a discrete Fourier transform sum
by an integral. The self-consistency equation
\begin{eqnarray}
s= \langle S(\vec{x}=0) \rangle = - \tanh[\beta \sum_{\vec{y}}
V(\vec{x}=0,\vec{y}) \langle S(\vec{y}) \rangle]
\end{eqnarray}
yields 
\begin{eqnarray}
\beta_{c}^{-1} = |\sum_{n=0}^{\infty} \frac{8(-1)^{n}}{(2n+1)\pi} v(\vec{k} = 
(2n+1) \vec{q})|.
\label{0}
\end{eqnarray}

The non-trivial solution (with a non-zero magnetization $s$) should be
self-consistent 
at all lattice sites (not only at $\vec{x} =0$). 
Self consistency at other values of $\vec{x}$ will lead to 
other lower bounds on $T_{c}$ which will read

\begin{eqnarray}
 |\sum_{n=0}^{\infty} \frac{8(-1)^{n}}{(2n+1)\pi}
v(\vec{k} = (2n+1) \vec{q})~  \cos [(2 n+1) \vec{q} \cdot \vec{x}|.
\label{x}
\end{eqnarray}

\section{$O(n \ge 2)$ ground states}
\label{n-gs}

Henceforth, the bulk of the article
focuses on continuous spin systems
having $n \ge 2$ components.
In an $O(n)$ spin system,
the spins $\{ \vec{S}(\vec{x}) \}$
have $n$ components and are all normalized to unity-
$\vec{S}^{2}(\vec{x})=1$.

A common intuitive 
picture held by many
is that in complex systems of 
the type presented
in Section(\ref{toy})
is the complex ground
states can arise. For instance,
in real systems, nanoscale phase separation
in the models of Section(\ref{toy})
can proceed through a certain class of 
viable ground states. Wigner crystals and arrays of dots 
are anticipated at large $q$. As $q$ is decreased, 
the spatial size of the dots increases, 
gradually leading to a state approaching macroscopic phase 
separation. Phase diagrams exhibiting devil staircases
are not uncommon in scalar systems. Albeit their intuitive
appeal and occurrence in many continuum
systems having scalar order parameters \cite{Seul},\cite{us},\cite{Low}, 
\cite{Bak},
ground states such as these do not 
generically occur in any $O(n \ge 2)$ system
on a cubic lattice. In the up and 
coming we will prove that, barring special
commensurability effects, any translationally
invariant $O(n \ge 2)$ system will generically
display only spiral (stripe like) or screw like \cite{nagamiya} 
ground states. For a discussion
of the occurrence of dots, Wigner crystals
etc. in the scalar case, the reader
is referred to \cite{Seul},\cite{us},\cite{Low},\cite{muratov}.
Configurations such as 
these were detected numerically (\cite{steve})
and analytically for the scalar Spherical model
(\cite{us}) for the spin models
that we discuss. We will explicitly prove,
for the first time, that 
these do not generically occur in incommensurate continuous
finite $n$ spin systems by deriving the relations
to be satisfied by the ground state configurations
containing a number of pairs of minimizing wave-numbers
$\{ \pm \vec{q}_{i}\}$. We will illustrate that special Wigner crystalline like
patterns may occur for special commensurate
wave-numbers alone. We explain how these
can arise in the spherical limit in 
Section(\ref{spherical}).

We first note that 
our previous ansatz $S(\vec{x}) = sign(\cos(\vec{q}_{1} \cdot \vec{x}))$
is readily fortified in the
O($n \ge 2$) scenario: here there is no need to
``digitize''- in  the spiral
\begin{eqnarray}
S_{1}(\vec{x}) = \cos (\vec{q} \cdot \vec{x}), ~
 S_{2}(\vec{x}) = \sin (\vec{q} \cdot \vec{x}), ~S_{i>2}(\vec{x}) = 0
\end{eqnarray}
the only non-zero Fourier components are
$\vec{S}(\vec{q}) , \vec{S}(-\vec{q})$.
In plain terms, this state can be constructed
with the minimizing wave-vectors only.
It follows that any ground state $g$ must be of
the form
\begin{eqnarray}
S_{i}(\vec{x})= \sum_{m} a_{i}^{m} \cos(\vec{q}_{m} \cdot \vec{x} +
 \phi_{i}^{m}). 
\end{eqnarray}
We now turn to the normalization of the spins at all sites, 
\begin{eqnarray}
1(\vec{x}) \equiv \sum_{i=1}^{n} S_{i}^{2}(\vec{x}) = \frac{1}{2} 
\sum_{m,m^{\prime}}
\sum_{i=1}^{n} a_{i}^{m} a_{i}^{m^{\prime}} 
\nonumber \\ (\cos(\phi_{i}^{m} + \phi_{i}^{m^{\prime}}) \cos[(\vec{q}_{m}+
\vec{q}_{m^{\prime}}) \cdot \vec{x}] \nonumber
\\ -\sin(\phi_{i}^{m}+\phi_{m^{\prime}})
\sin[(\vec{q}_{m}+\vec{q}_{m^{\prime}}) \cdot \vec{x}]  
\nonumber
\\ + \cos (\phi_{i}^{m} -\phi_{i}^{m^{\prime}}) \cos[(\vec{q}_{m} -
\vec{q}_{m^{\prime}}) \cdot \vec{x}] \nonumber 
\\ -\sin(\phi_{i}^{m}-\phi_{i}^{m^{\prime}})
\sin[(\vec{q}_{m} -\vec{q}_{m^{\prime}}) \cdot \vec{x}]). 
\end{eqnarray}
If $1(\vec{x})= 1$ is to hold identically for all sites $\vec{x}$,
 then all non-zero Fourier 
components must vanish. 
For the $\{ \cos(\vec{A} \cdot \vec{x})  \}$  Fourier components:
\begin{eqnarray}
0= [\sum_{\vec{q}_{m}+\vec{q}_{m^{\prime}}= \vec{A}}~~
 \sum_{i=1}^{n} a_{i}^{m} a_{i}^{m^{\prime}}
\cos (\phi_{i}^{m} + \phi_{i}^{m^{\prime}}) \nonumber \\
+ \sum_{\vec{q}_{m}-\vec{q}_{m^{\prime}} = \vec{A}}~~ \sum_{i=1}^{n} a_{i}^{m}
a_{i}^{m^{\prime}} \cos(\phi_{i}^{m}-\phi_{i}^{m^{\prime}})],
\end{eqnarray}
and a similar relation is to be satisfied by the $\{ \sin(\vec{A} \cdot
 \vec{x}) \}$ components:
\begin{eqnarray}
0= [\sum_{\vec{q}_{m}+\vec{q}_{m^{\prime}}= \vec{A}}~~ \sum_{i=1}^{n}
 a_{i}^{m} a_{i}^{m^{\prime}}
\sin (\phi_{i}^{m} + \phi_{i}^{m^{\prime}}) \nonumber \\
+ \sum_{\vec{q}_{m}-\vec{q}_{m^{\prime}} = \vec{A}}~~ \sum_{i=1}^{n}
a_{i}^{m} a_{i}^{m^{\prime}} \sin (\phi_{i}^{m}-\phi_{i}^{m^{\prime}})]
\end{eqnarray}
\begin{eqnarray}
a_{i}^{m} \cos \phi_{i}^{m} \equiv v_{i}^{m},~~ a_{i}^{m} \sin 
\phi_{i}^{m} \equiv u_{i}^{m}.
\end{eqnarray} 
The $\{ \cos(\vec{A} \cdot \vec{x}) \}$ and $\{ \sin(\vec{A} \cdot
 \vec{x}) \}$ conditions read 
\begin{eqnarray}
0= \sum_{\vec{q}_{m}+\vec{q}_{m^{\prime}}= \vec{A}}~~ \sum_{i=1}^{n}
[v_{i}^{m} v_{i}^{m^{\prime}} - u_{i}^{m} u_{i}^{m^{\prime}}]
\nonumber
\\ +\sum_{\vec{q}_{m}-\vec{q}_{m^{\prime}} = \vec{A}}  \sum_{i=1}^{n} 
[v_{i}^{m} v_{i}^{m^{\prime}} + u_{i}^{m} u_{i}^{m^{\prime}}] \nonumber
\\ 0= \sum_{\vec{q}_{m}+\vec{q}_{m^{\prime}}= \vec{A}}~~ \sum_{i=1}^{n} 
[v_{i}^{m} u_{i}^{m^{\prime}}+u_{i}^{m} v_{i}^{m^{\prime}}]\nonumber
\cr + \sum_{\vec{q}_{m}-\vec{q}_{m^{\prime} = \vec{A}}} ~~ \sum_{i=1}^{n}
[v_{i}^{m} u_{i}^{m^{\prime}} - u_{i}^{m} v_{i}^{m^{\prime}}] 
\end{eqnarray} 
Let us now consider the case of two
pairs of minimizing modes. For two 
pairs of wave-vectors $\pm \vec{q}_{1}$ and
$\pm \vec{q}_{2}$ minimizing
$v(\vec{k})$, both not equal half a reciprocal lattice vector:
$(0,0,..,), (\pi,0,...,0), ~(0,\pi,0,...,0)$,...,
$(\pi,\pi,0,...,0),...,(\pi, \pi,...,\pi)$,
the vector $\vec{A}$ (up to an irrelevant sign) may attain 
four non-zero values:
$\vec{A} = 2\vec{q}_{1},2\vec{q}_{2},\vec{q}_{1} \pm \vec{q}_{2}$.

When $\vec{A} = \vec{q}_{1} + \vec{q}_{2}$, the conditions are
\begin{eqnarray}
0 = \sum_{i=1}^{n} [v_{i}^{1} v_{i}^{2} - u_{i}^{1} u_{i}^{2}], 
~ 0= \sum_{i=1}^{n} [v_{i}^{1} u_{i}^{2} + u_{i}^{1} v_{i}^{2}].
\end{eqnarray} 
When $\vec{A} = \vec{q}_{1} - \vec{q}_{2}$, these conditions read
\begin{eqnarray}
0= \sum_{i=1}^{n} [v_{i}^{1} v_{i}^{2} + u_{i}^{1} u_{i}^{2}], 
~0 = \sum_{i=1}^{n} [v_{i}^{1} u_{i}^{2} - u_{i}^{1} v_{i}^{2}]
\end{eqnarray}
For $\vec{A} = 2\vec{q}_{\alpha}$ ~($\alpha =1,2$):
\begin{eqnarray}
0= \sum_{i=1}^{n} [v_{i}^{\alpha} v_{i}^{\alpha} - u_{i}^{\alpha}
 u_{i}^{\alpha}],
0= 2 \sum_{i=1}^{n} u_{i}^{\alpha} v_{i}^{\alpha} 
\end{eqnarray}
Next, we define 
\begin{eqnarray}
\vec{U}^{\alpha} \equiv (u_{i=1}^{\alpha},u_{2}^{\alpha},...,u_{n}^{\alpha})
 \nonumber
\\ \vec{V}^{\alpha} \equiv (v_{1}^{\alpha},...,v_{n}^{\alpha})
\end{eqnarray}
The previous conditions for  $\vec{A} = \vec{q}_{1} \pm \vec{q}_{2}$, 
$2q_{1,2}$ imply that 
\begin{eqnarray}
\vec{V}^{1} \cdot \vec{U}^{2} = \vec{U}^{1} \cdot \vec{V}^{2} =0 \nonumber
\\ \vec{V}^{1} \cdot \vec{V}^{2} = \vec{U}^{1} \cdot \vec{U}^{2} =0 \nonumber
\\ \vec{U}^{1} \cdot \vec{V}^{1} = \vec{U}^{2} \cdot \vec{V}^{2} =0. 
\label{ortho}
\end{eqnarray}
The four vectors $\{ \vec{U}^{1},\vec{U}^{2},\vec{V}^{1},\vec{V}^{2} \}$
are all mutually orthogonal.  The number of spin components $n \ge 4$.
Two additional demands that follow are
\begin{eqnarray}
\vec{V}^{\alpha} \cdot \vec{V}^{\alpha} = \vec{U}^{\alpha} \cdot
 \vec{U}^{\alpha} \nonumber
\\ \sum_{\alpha=1}^{2} [\vec{V}^{\alpha} \cdot \vec{V}^{\alpha} +
 \vec{U}^{\alpha} \cdot \vec{U}^{\alpha}] = 2 \sum_{\alpha} \vec{V}^{\alpha}
 \cdot \vec{V}^{\alpha} = 2.
\label{norm}
\end{eqnarray}
The last equation is the normalization condition- the statement that the
 coefficient
of $\cos (\vec{A} \cdot \vec{x})$, when $\vec{A} =0$, is equal to 1. 
For the case of a single pair of wave-vectors, $\pm \vec{q}_{1}$, 
~~$\vec{A} = 2 \vec{q}_{1}, 0$ and the sole conditions are encapsulated in
the last of equations(\ref{ortho}) and in equation (\ref{norm}).
A moment's reflection reveals that this only allows for a spiral in the plane
 defined
by $\vec{U}^{1}$ and $\vec{V}^{1}$.
When $n<4$ there are no configurations  
which  satisfy $\vec{S}^{2}(\vec{x}) =1$
identically for all sites $\vec{x}$ [excusing those having
$2(\vec{q}_{i}+\vec{q}_{j})=\vec{A}$ is equal to
a reciprocal lattice vector]
that are a superposition of exactly
two modes. For instance, with the convention
that a ``double checkerboard state'' 
is a  
Neel state of $2 \times 2$ 
blocks \cite{Low} ($p_{1}=p_{2}=2$
in the two dimensional planar
version of Eq.(\ref{ppp}),
we note that 
a (double checkerboard state along 
the $i=1~ axis$) $\otimes$  (a spiral in the $23~~ plane$) 
has pairs $(i,j)$ is a configuration 
in which $\vec{A} = 2(\vec{q}_{i} +\vec{q}_{j})$
is a reciprocal lattice vector.
As the number of minimizing modes $\{ \vec{q}_{m} \}$ increases, some of
the conditions may degenerate into one, e.g.  if 
$(\vec{q}_{1}+ \vec{q}_{2}) = (\vec{q}_{3}-\vec{q}_{2})$
(i.e. the modes are collinear). This degeneracy is
the a second route that might allow for Ising configurations 
which are superpositions of several ``good''minimum energy 
modes $\exp(i \vec{q} \cdot \vec{x})$. 
The highly degenerate Ising ground states that
we have constructed previously can fall under
either one of these categories.
If neither one of these situations occurs, Ising states
cannot be superpositions of several minimum energy modes:
we will be left with too many equations of 
constraints with too few degrees of freedom.
For three pairs of minimizing modes, none of which is half
a reciprocal lattice vector,  $\{ \pm \vec{q}_{m} \}_{m=1}^{3}$ 
with
\begin{eqnarray}
\vec{q}_{w} \pm \vec{q}_{t} \neq \vec{q}_{r} \pm \vec{q}_{s} \neq 2 \vec{q}_{w}
\end{eqnarray}
for all $w \neq t$, ~ and $r \neq s$, 
conditions similar to those that previously written for
 $\vec{A} = \vec{q}_{1} \pm \vec{q}_{2}$, now hold for all 
$(\vec{q}_{w} \pm \vec{q}_{t})$.
\begin{eqnarray}
\vec{U}_{\alpha} \cdot \vec{U}_{\beta} = \vec{U}^{2}_{\alpha}
 ~\delta_{\alpha,\beta} \nonumber
\\ \vec{V}_{\alpha} \cdot \vec{V}_{\beta} = \vec{V}^{2}_{\alpha}
 ~\delta_{\alpha,\beta} \nonumber
\\ \vec{U}_{\alpha} \cdot \vec{V}_{\beta} =0
\end{eqnarray}
The relation $\vec{U}_{\alpha} \cdot \vec{V}_{\alpha}=0$
 ($\alpha = \beta$ in the last Eq above) is 
enforced by setting $\vec{A} = 2 \vec{q}_{\alpha}$.
Thus, when exactly three pairs of minimizing wave-vectors 
satisfying the equation are present,
the vectors $\{ \vec{U}_{\alpha},\vec{V}_{\alpha} \}$ define 
a 6-dimensional space, and hence $n \ge 6$. For $p$ pairs of minimizing
wave-vectors, $n$ must be at least $2p-$dimensional.
This bound is saturated when $\vec{S}$ is (a spiral state in the $12-plane$)
$\otimes$ (a spiral in the $34-plane$) $\otimes ... \otimes$ (a spiral in the
$2p-1,2p~~ plane$), i.e. 
\begin{eqnarray}
(a_{1} \cos(\vec{q}_{1} \cdot \vec{x}+ \phi_{1}), a_{1} \sin(\vec{q}
 \cdot \vec{x} + \phi_{1}),... \nonumber
\\ ,a_{p} \cos(\vec{q}_{p} \cdot \vec{x} + \phi_{p}),
a_{p} \sin(\vec{q}_{p} \cdot \vec{x} + \phi_{p}))
\end{eqnarray}
with $\sum_{\alpha=1}^{p} a_{\alpha}^{2} =1$.
When wave-vectors with $\vec{q}_{w} \pm \vec{q}_{t} = \vec{q}_{r} \pm
\vec{q}_{s}$ or $\vec{q}_{w} \pm \vec{q}_{t} = 2 \vec{q}_{r}$ are present,
pairs of conditions degenerate into single linear 
combinations. Thus far we assumed that for all $i$ and $j$, 
$\vec{A} = 2(\vec{q}_{i}+ \vec{q}_{j})$ is not a reciprocal lattice 
vector, s.t. $\sin(\vec{A} \cdot \vec{x})$ is not identically
zero at all $\vec{x} \in Z^{d}$. 
We term the such a $p=2$ configuration a bi-spiral. It is
simple to see by counting the number of degrees of freedom for
$n=4$, that the bi-spirals overwhelm states having only
one mode $\pm \vec{q}_{1}$. This is a  simple instance of
a general trend: High $p$ states are statistically preferred.
Moreover, as we shall see later, they are more stable
against thermal fluctuations.

In summary, we outlined a way to 
determine all $O(n\ge 2)$ 
ground states, whether commensurate 
or incommensurate, for a given kernel
$V(\vec{x},\vec{y}) = V(\vec{x}-\vec{y})$.
Whenever $n \ge 2$, any ground state configuration can be decomposed
into Fourier components,
${\vec S}^{g}({\vec x})= \sum_{i=1}^{|{\cal{M}}|} \left \{ 
\cos[{\vec q}_{i} \cdot \vec x] + {\vec b}_i\sin[{\vec q}_{i}
\cdot \vec x]
\right \}$. The vectors ${\vec q}_{i}$ are chosen from the set of wave vectors
which minimize $v(\vec k)$. Here, the modulus 
$|{\cal{M}}|$ is the number of minimizing modes 
(the ``measure'' of the modes on the minimizing
surface $M$. As long as these wave-vectors ${\vec q}_{i}$
which minimize $v(\vec k)$ 
are ``non-degenerate'', in the sense
that the sum of any pair of wave vectors,  
$\vec{q}_{i} \pm \vec{q}_{j}$  is not equal to
the sum of any other pair of wave vectors, and ``incommensurate'' in 
the sense that for all $i$ and $j$, 
$2(\vec{q}_{i}+\vec{q}_{j})$ is not equal to
a reciprocal lattice vector, then the 
that the condition $[\vec{S}^{g}(\vec{x})]^{2}=1$
can be satisfied only if $|{\cal{M}}| \le n/2$.  (In our toy model 
of the doped Mott insulator, these conditions 
are always satisfied for $Q < 4$.)
Thus, for $n\le 3$ 
only simple spiral ($|{\cal{M}}|=1$) ground-states are permitted, while for
$n=4$, a double spiral saturates the bound.
Thus, generically, for $2 \le n<4$ all
ground states will be spirals
containing only one mode. 
The reader should bear in mind that 
in the usual short range ferromagnetic
case, the ground states
are globally  $SO(n)$ symmetric and are
labeled by only $(n-1)$ continuous
parameters. 
Here, for each minimizing mode there
are $(2n-3)$ continuous internal 
degrees of freedom labeling 
all possible spiral ground states.
For $n>2$  this guarantees a much
higher degeneracy than that
of the usual ferromagnetic
ground state. 
If there are many minimizing modes 
(e.g. if the minimizing manifold $M$
were endowed with $SO(d-1)$ symmetry)
then the ground state degeneracy
is even larger! When $n\ge 4$,  there are (generically)
even many more ground states
(poly-spirals). These poly-spiral states have a degeneracies
larger than those of simple 
spiral. Their degeneracy
\begin{eqnarray}
g= p(2n-2p-1)|M|^{p},
\end{eqnarray}
where $|M|$ is the number
of minimizing modes. 
We just proved
that if frustrating interactions
cause the ground states to be modulated
then the associated  ground state
degeneracy (for $n>2$) is much
larger by comparison to the 
usual ferromagnetic ground states.

\section{Spin Stiffness} 
\label{s-stiffness}

In this section, we focus for the sake of concreteness alone,  
on the Coulomb Frustrated Ferromagnet.
A similar analysis, with different results, 
may be reproduced for all other interactions.
When $Q>0$ in the in the continuum (small k limit) 
of the Coulomb Frustrated Ferromagnet
of Eq.(\ref{Q}), with all $O(k^{4})$  and higher terms 
ignored, the minimizing
modes of $v_{Q}(\vec{k})$ 
lie on the surface of a sphere
 $\{ M_{Q}: \vec{q}^{2} = \sqrt{Q}\}$. 
As $Q \rightarrow 0$, this surface $M_{Q}$ 
shrinks and shrinks yet is still a $(d-1)$ 
dimensional surface of a sphere. When $Q=0$, 
the minimizing manifold evaporates into
 a single point $\vec{q} =0$. This sudden change 
in the dimensionality has profound consequences. 
As we shall see shortly, it lends itself to suggest
(quite strongly) that order is inhibited for a
Heisenberg ($n=3$) realization of our model.
Before doing so, let
us indeed convince ourselves,
on an intuitive level, that the large
degeneracy in $\vec{k}-$ space
brought about by the frustration
gives rise to a reduced spin stiffness.
We now assume an ordered spiral state 
of momentum $\vec{q}$ on a cubic lattice
(after all, we proved that in the general 
incommensurate case, these are the only
ground states), 
and we examine the energy cost
of varying the modulation wave-vector in a direction
parallel and transverse to $\vec{q}$.
The exact origin of the spiral ground is
completely irrelevant (whether it is the 
the Coulomb Frustrated Model, the membrane
model or other very different models)-
all that matters is that the interaction
kernel attains its minima at finite
incommensurate momenta $|\vec{q}|$ which
cover a spherical shell of radius $q$. 
For general rotationally symmetric incommensurate systems, 
$Q$ may be regarded as a convenient shorthand, $Q=q^{4}$.
The upshot of the below, rather trivial,
calculation is that rotationally invariant incommensurate 
continuous spin systems are very susceptible 
to transverse perturbations- a transverse
twist may incur no energy penalty
in the thermodynamic limit.

\subsection{Longitudinal Spin Stiffness}
We first examine the energy cost of a longitudinal twist to
find that for highly incommensurate frustrated systems with
shells of minimizing modes, the energy cost for
such a perturbation is just the same as in a nearest
neighbor three dimensional XY ferromagnet. In a longitudinal
twist  
\begin{eqnarray}
\vec{S}(\vec{x}) = \cos( \frac{2 \pi}{L} x + q x) \hat{e}_{1} + 
\sin(\frac{2 \pi}{L} x + qx ) \hat{e}_{2} + \delta \vec{S} 
\nonumber 
\\ = \hat{e}_{1} \cos(\vec{k} \cdot \vec{x}) + 
\hat{e}_{2} \sin (\vec{k} \cdot \vec{x}) + \delta \vec{S}
\end{eqnarray}   
with $\vec{k} = (\frac{2 \pi}{L}+ q ) \hat{e}_{1}$.
The energy cost of this twisted state relative to the ground state is
\begin{eqnarray}
\Delta E[\{ \vec{S}(\vec{x}) \} ] = \frac{1}{2N} \sum_{\vec{k}^{\prime}} 
[v(\vec{k}^{\prime})-v(\vec{q})] |\vec{S}(\vec{k}^{\prime})|^{2}
\end{eqnarray}
Ignoring $\delta \vec{S}$ contributions,
\begin{eqnarray}
\Delta E = \frac{N}{2}[v(\vec{k})-v(\vec{q})] = \frac{N}{2 \sqrt{Q}} 
[( \frac{2 \pi}{L} + q)^{2} - q^{2}]^{2} \approx \frac{8 \pi^{2}N}{L^{2}}. 
\nonumber
\end{eqnarray}
It is readily seen that this energy gain exactly coincides
with the energy gain incurred by a uniform longitudinal twist in 
the three dimensional  
nearest neighbor ferromagnetic $XY$ system, $ \vec{S}(\vec{x}) 
= \cos [ \frac{2 \pi x}{L}] \hat{e}_{1} + \sin 
[\frac{2 \pi x}{L}] \hat{e}_{2}$.

\subsection{Transverse Spin Stiffness}
When subjected to a transverse twist, the system
responds with only a quartic restoring potential. 
For a transverse twist, 
\begin{eqnarray}
\vec{S}(\vec{x}) = \cos[\frac{2 \pi x}{L} + qy] \hat{e}_{1} +
 \sin[\frac{2 \pi x}{L} + qy] \hat{e}_{2} + \delta \vec{S}
\end{eqnarray}
we find that as the difference of
the energy kernels,  $v(\vec{k}= \vec{q} \hat{e}_{2} 
+ \frac{2 \pi}{L} \hat{e}_{1}) - v(\vec{q}) = 
\frac{16 \pi^{4}}{L^{4} \sqrt{Q}}$,
the energy gain, ignoring $\delta \vec{S}$, is 
$\Delta E = \frac{8 \pi^{4} N}{L^{4} \sqrt{Q}}$. 
This energy penalty for
a transverse twist vanishes 
as $L \rightarrow \infty$ in $d=3$- there is a complete loss of stiffness
against transverse fluctuations. As $N = L^{d}$, this energy penalty
scales as ${\cal{O}}(\frac{1}{L})$. We note, in passing, 
that, to this order, this energy penalty matches with that of
an XY chain, 
\begin{eqnarray}
\Delta E \sim \frac{2 \pi^{2}}{L^{2}} N = \frac{2 \pi^{2}}{L} = 
{\cal{O}}(\frac{1}{L}).
\end{eqnarray}
An effective reduction in dimensionality seems
to occur- the system effectively behaves
as a one dimensional system when exposed to transverse
perturbations.

In the general case, for all
incommensurate systems,
this simple spin wave analysis
yields a response to a twist
$\vec{\delta}$ with an effective
kernel $E_{low}(\vec{\delta}) = v(\vec{q} + \vec{\delta}) - v(\vec{q})$.
For systems with a spherical shell of incommensurate minimizing modes,
$E_{low} \simeq A_{\perp} \delta_{\perp}^{4}$ 
for momentum deviations transverse to the minimizing
$\vec{q}$, while $E_{low} \simeq  A_{||} \delta_{||}^{2}$
for deviations parallel to $\vec{q}$.  Such a fluctuation
spectrum is reminiscent to that obtained in smectic 
liquid crystals. It will be noted, however, that at this
stage of approximation, $E_{low} \neq  
 A_{\perp} \delta_{\perp}^{4} +  A_{||} \delta_{||}^{2}$,
for general deviations $\vec{\delta}$: the precise fluctuation
spectrum is much softer than that of a smectic 
liquid crystal (e.g. $E_{low}$ vanishes for  $\vec{\delta}$
connecting $\vec{q}$ with another vector 
on the minimizing manifold). 

\section{Thermal fluctuations of any 
translationally invariant XY model}
\label{XY-fluct}

In the previous section, we examined the 
response of an arbitrary incommensurate system
having a shell of minimizing modes 
to various twists by simply examining
the energy difference between a ground
state spiral and a spiral of another
wave-number. We found that the
system was very unstable to 
transverse twists. We now
go one step further and 
completely treat the ``soft-spin'' version of the
classical XY model with arbitrary incommensuration
(not only those with a shell of minimizing modes). 
Towards the end of this section, we investigate 
what our general results imply in the case of rotationally symmetric 
incommensurate systems (which have a shell of minimizing modes) 
where we will find a
link to smectic liquid crystal thermodynamics. 

In this general theory, 
we include the non-linear interaction $H_{1}$, 
\begin{eqnarray}
H_{soft} = H_{0} +  u \sum_{{\vec  x}} [\vec{S}^{2}({\vec  x})-1]^{2}
 \equiv H_{0} + H_{1}
\end{eqnarray}
with small $u>0$,
and forget about the normalization
conditions $|\vec{S}({\vec  x})|=1$
at all lattice sites $\vec{x}$.
(The normalized, ``hard-spin'', version can be viewed as the 
$u\rightarrow \infty$ limit of the soft-spin model.)
As we proved previously, the only generic ground 
states (for both hard- and soft-spin models)
when the spins $\vec{S}(\vec{x})$ have 
two (and also three) components are spirals
\begin{eqnarray}
S_{1}^{ground-state}(\vec{x}) = \cos(\vec{q} \cdot \vec{x}); 
~ S_{2}^{ground-state}(\vec{x}) = \sin(\vec{q} \cdot \vec{x}). 
\nonumber
\end{eqnarray}
These are also the lowest energy eigenstates
of $H_{soft}$. We will shortly expand $H_{soft}$ about these
ground states, keeping only the
lowest order (quadratic) terms
in the fluctuations $\delta S$.
The quadratic term in $\{ \delta S_{i}(\vec{k}) \}$ stemming
from $H_{soft}$
is the bilinear $\frac{u}{N} (\delta S)^{+} M (\delta S)$
where 
\begin{eqnarray}
(\delta S)^{+} = (\delta S_{1}(-\vec{k}_{1}), \delta S_{2}(-\vec{k}_{1}), 
\delta S_{1}(-\vec{k}_{2}) \delta S_{2}(-\vec{k}_{2}), \nonumber
\\
\delta S_{1}(-\vec{k}_{3}), \delta S_{2}(-\vec{k}_{3}), ..., 
\delta S_{1}(-\vec{k}_{N}), \delta S_{2}(-\vec{k}_{N}))
\end{eqnarray}
and the matrix $M$ 
reads 
$$
\pmatrix{ 4 & 0 & . & .& 1 & i & . & . & . & . \cr
 0 & 4 & . & . & i &-1 & . & . & . & .\cr 
. & . & . & . & . & . & . & .  & . & . \cr
. & . & . & . & . & . & . & . & . & . \cr
 1 & -i & . & . & 4 & 0 & . & . & 1 & i \cr
 -i & -1 & . & . & 0 & 4 & . & . & i & -1 \cr
 . & . & . & . & . & . & . & . & . & . \cr
 . & . & . & . & . & . & . & . & . & . \cr
. & . & . & . & 1 & -i & . & . & 4 & 0 \cr
 . & . & . & . & -i &-1 & . & . & 0 & 4}.
$$
The sub-matrices are $(2 \times 2)$ matrices in the internal spin indices.
The off diagonal blocks are separated from the diagonal ones
by wave-vectors $(\pm 2 \vec{q})$. 
Note that $\langle \vec{k}|M|\vec{k}^{\prime} \rangle = 
M(\vec{k} - \vec{k}^{\prime})$.  
Making a unitary (symmetric Fourier) 
transformation to the real space basis:
$|\vec{x} \rangle \equiv N^{-1/2} \sum_{\vec{k}} e^{i \vec{k} \cdot
\vec{x}} |\vec{k} \rangle$,
the matrix $M$ becomes block diagonal, $
\langle \vec{x}| M|\vec{x}^{\prime} \rangle = \hat{M}(\vec{x})
\delta_{\vec{x}, \vec{x}^{\prime}}$.
Diagonalizing in the internal spin basis, we find that
\begin{equation}
\lambda_{\pm} = 6,2.
\label{sep}
\end{equation}
These eigenvalues may be regarded, in the usual
ferromagnetic case (the limit $\vec{q} = 0$) as 
a two step (state) potential barrier separating 
the two polarizations. I.e., the normalization 
constraint of the XY spins (embodied in $H_{1}$) 
gives rise to an effective binding
interaction. As we shall later see, 
when the number of spin components $n$
is odd, one spin component will
remain unpaired. $H_{1}$ literally 
``couples'' the spin polarizations.
Employing Eq.(\ref{sep}), the corrected 
fluctuation  spectrum $\{ \psi_{m} \}_{m=1}^{N}$
(to quadratic order)
satisfies a Dirac like equation
\begin{eqnarray}
[U^{+}~v(-i \partial_{x})~U + 2 u ~ \pmatrix{ 2 & 0 \cr
0 & 6}~ ]~ U^{+}|\psi_{m}(\vec{x}) \rangle \nonumber
\\ = E_{m}~ U^{+}|\psi_{m}(\vec{x}) \rangle.
\end{eqnarray}
Here, 
\begin{eqnarray}
U = \pmatrix{ \frac{\sin (2 \vec{q} \cdot \vec{x})}{2
 \cos(\vec{q} \cdot \vec{x})} & \frac{1}{2} \cr
\frac{-1 - \cos(2 \vec{q} \cdot \vec{x})}{2 \cos(\vec{q} \cdot \vec{x})} & 
\frac{1-\cos(2 \vec{q} \cdot \vec{x})}{2 \sin (\vec{q} \cdot \vec{x})}}.
\end{eqnarray}
Alternatively, expanding in the 
fluctuations $\delta S(\vec{k})$,
leads to bilinear 
$(\delta S)^{+} {\cal{H}} (\delta S)$
where   
\begin{eqnarray}
{\cal{H}}_{\vec{k},\vec{k}}= \pmatrix{v(\vec{k})+ 8u & 0\cr
0 & v(\vec{k})+ 8u }
\label{HKK}
\end{eqnarray}
along the diagonal, and ${\cal{H}}_{\vec{k},\vec{k} 
\pm 2\vec{q}} = 2u(\sigma_{3} \pm i \sigma_{1})$
off the diagonal. Next, we perform a unitary transformation
$\exp(i \frac{\pi}{4} \sigma_{1}) {\cal{H}}_{\vec{k},\vec{k} \pm 2 \vec{q}}
\exp(- i \frac{\pi}{4} \sigma_{1}) = 2u \sigma^{\pm}$, 
while ${\cal{H}}_{\vec{k},\vec{k}}$ of Eq.(\ref{HKK}) is unchanged. 
Till now, all that we stated, held for arbitrarily
large $u$- our only error was neglecting 
$O((\delta S)^{3})$ terms by comparison 
to $O((\delta S)^{2})$. Note that the main difficulty 
with the approach taken till now was the coupling
between $\vec{k}$ and $\vec{k} \pm 2 \vec{q}$:
i.e. $\vec{k}$ is coupled to 
$\vec{k} \pm 2 \vec{q}$, while $\vec{k}+ 2 \vec{q}$
is coupled to $\vec{k} + 4\vec{q}$ and $\vec{k}$, and so on. 
Unless $\vec{q}$ is of low commensurability
an exact solution to this problem is impossible.
Equations with similar structure appear in very different arenas 
(e.g. two dimensional Bloch electrons in a magnetic
field \cite{hofs}, \cite{azbel}) 
and such systems are often addressed via a continued fraction 
representation or by mapping the problem onto a 
Harper like Equation. To make rapid progress and to elucidate
the similarity between incommensurate systems
and smectic liquid crystals, here we will address our
problem is a very direct and short fashion. Let us assume that $u$ is small. 
In this case the lowest eigenstates of the fluctuation
matrix will contain only a superposition of the low lying $\vec{k}$- states
(i.e. those close to the ($d-1)$ dimensional $M$ ($q>0$)). 
If $\vec{k}_{1} = \vec{q}+ \vec{\delta}$ is close to $M$, 
then the only important modes in the sequence 
$\{S_{i}(\vec{k} =  \vec{k}_{1}+ 2n\vec{q})\}$ are $\vec{k}_{1}$, and
$\vec{k}_{2}=\vec{k}_{1} - 2 \vec{q} = -\vec{q}+\vec{\delta}$. 
The sub-matrix in the 
relevant sector reads
\begin{eqnarray}
\pmatrix{ v(\vec{k}_{1})+8u & 0 & 0 & 4u \cr 
0 & v(\vec{k}_{1})+8u & 0 & 0 \cr
0 & 0 & v(\vec{k}_{2})+8u & 0 \cr
4u & 0 & 0 & 0 & v(\vec{k}_{2})+8u}.
\nonumber
\end{eqnarray}
The lowest eigenvalue reads
\begin{eqnarray}
E_{low} = \frac{1}{2}[v(\vec{k}_{1})+v(\vec{k}_{2})]+4u \nonumber
\\ -\frac{1}{2}
 \sqrt{[v(\vec{k}_{1})-v(\vec{k}_{2})]^{2}+64 u^{2}}.
\label{spec}
\end{eqnarray}
Equivalently, this can be determined from the direct computation
of the determinant to $O(u^{2})$: to obtain $O(u^{2})$ contributions
we need to swerve off the diagonal twice. 
\begin{eqnarray*}
\! \! \! \! \lefteqn{\! \! \! \! \! \! \! \!
\det{\cal{H}} = \prod_{i=1}^{N} [v(\vec{k}_{i})+8u]^{2} 
-(4u)^{2} \sum_{j}~ [v(\vec{k}_{j})+ 8u] } \\
& &  \times[v(\vec{k}_{j}+2 \vec{q})+8u]~ \prod_{\vec{k}_{i}
 \neq \vec{k}_{j},\vec{k}_{j} + 2 \vec{q}}
[v(\vec{k}_{i})+8u]^{2}  \\
\end{eqnarray*}
The fluctuation spectrum is trivially determined
by replacing $v(\vec{k})$ by $[v(\vec{k}) - E]$ in $\det{\cal{H}}$
and setting it to zero. 
To this order we re-derive $E_{low}$. 
Higher order terms in the determinant may be trivially
computed. The partition function is $Z = const ~[\det {\cal{H}}]^{-1/2}$.
For any $u$, no matter how small, there exists a neighborhood
of wave-vectors $\vec{k}$ near $\vec{q}$ 
such that $|v(\vec{k})-v(\vec{k}+ 2 \vec{q})| \ll u$
and as before we may re-expand the characteristic equation
for these low lying modes, solve a simple 
quadratic equation, expand in the components
of $(\vec{k}-\vec{q})$ and obtain a simple dispersion
relation.

Till now, our results, held for arbitrary incommensurate
XY systems. We now investigate what transpires
when the minimizing modes
form a continuous shell M
(e.g. all incommensurate systems
with a rotational symmetry
that secures, as $q \neq 0$, a shell of 
radius q of minimizing modes about 
the origin). For concreteness, if 
$\vec{\delta} = \vec{\delta}_{\perp} + \delta_{||}~ \hat{e}_{||}$,
with $\perp~,~||$ denoting directions orthogonal, and parallel 
to $\hat{n} \perp M$, 
then  by expanding Eq.(\ref{spec}) to lowest orders for the
example presented in Eq.(\ref{Q})
\begin{eqnarray}
E_{low} = A_{\perp}~ \delta_{\perp}^{4}+ A_{||}~ \delta_{||}^{2},
\label{twist*}
\end{eqnarray}
where
\begin{eqnarray}
A_{||} = \frac{1}{2} \frac{d^{2}v(|\vec{k}|)}{dk^{2}}|_{|k|=q};
 ~A_{\perp}= \frac{A_{||}}{4q^{2}}.
\label{AAp}
\end{eqnarray}
Examining the fluctuations about
an ordered state, we find 
that the effect of thermal
fluctuations is reduced
by comparison to 
the dispersion
relation in the 
large $n$ model
and spin wave stiffness
analysis. Nonetheless, at this
level of analysis,
order is still inhibited,
much unlike many models
wherein ``order out of
disorder'' occurs  \cite{Villain}
by entropic stabilization
about certain viable 
ground states. This dispersion is 
akin to the fluctuation
spectrum of the smectic liquid crystals 
\cite{Nielsen} which is well known to give rise 
to algebraic decay of correlations
at low temperatures.
In our case,  $\vec{q} \neq 0$ and 
thus the correlations
should have an oscillatory prefactor. 
To be more concrete we find that,
for general rotationally symmetric 
incommensurate spin systems having 
a minimizing shell of modes at $|\vec{q}|=q>0$, 
the correlator (in cylindrical 
coordinates) is
\begin{eqnarray}
G(\vec{x}) \equiv \langle \vec{S}(0) \cdot \vec{S}(\vec{x}) \rangle 
\simeq \frac{4 d^{2}}{x_{\perp}^{2}}~ \exp[- 2 \eta \gamma -
\eta E_{1}(\frac{x_{\perp}^{2} q}{2 x_{||}})]
\nonumber \\ \times ~ \cos[qx_{||}],
\label{ALRO}
\end{eqnarray}
where $\eta= \frac{k_{B}T}{16 \pi}q$, $d= \frac{2 \pi}{\Lambda}$ 
where $\Lambda$ is the  
ultra violet momentum cutoff, $\gamma$ is Euler's const., 
\begin{eqnarray}
E_{1}(z) = -\gamma - \ln z - \sum_{n=1}^{\infty} (-)^{n} \frac{z^{n}}{n(n!)}
\end{eqnarray}
is the exponential integral.
For the uninitiated reader, 
we present this standard derivation in
the appendix. 

The thermal fluctuations $\int d^{d}k / E_{low}(\vec{k})$ 
diverge as $[- \ln |\epsilon|]$ with a
$\epsilon$ a lower cutoff on $|\vec{k}-\vec{q}|$ \cite{dive}.
Such a logarithmic divergence is also
encountered in ferromagnetic two dimensional 
$O(2)$ system if it were exposed to
the same analysis. Indeed, both 
models share similar characteristics
albeit having different physical 
dimensionality. Thus far, 
our analysis was performed on
incommensurate structures
with minimizing shells in $\vec{k}$-space.
For frustrated systems having special
commensurabilities, intricacies 
may arise (some are explicitly detailed 
in \cite{intracate_comm}).

{\bf{Relations 
to Smectic Liquid Crystals}}

We now fortify our earlier conclusion
regarding a tight relation between
incommensurate continuous spin systems
and smectic liquid crystals. For 
incommensurate spin systems
with minima on a 
$(d-1)$ dimensional shell
in $\vec{k}$ space,
the resulting spectra
attained by both 
the Dirac like equation
and the simple more naive
spin stiffness analysis,
matches that of
smectic liquid
crystals. 
Smectic ordering involves a breaking of
rotational symmetry and of translational
symmetry along one direction
alone. Let $\tilde{u}$ 
be the displacement of the smectic along
the z-direction.
When the wave-vector $\vec{k}$ is along the z-direction
the displacement is longitudinal, and the energy is of the 
elastic form  $\frac{1}{2}B k_{||}^{2} |\tilde{u}(\vec{k})|^{2}$,
where $B$ is the compressibility for the smectic layers. 
When $\vec{k}$ is normal to z, the displacement is
transverse to the layer separation. No second order in
$\vec{k}_{\perp}$ the displacement costs no energy. 
Here, the restoring force is associated with a director splay 
distortion leading to an elastic energy 
$\frac{K}{2} (\vec{\nabla} \cdot \vec{n})^{2}$
with $K$ the splay constant. In, e.g. Smectic A, 
as $\hat{n}$ is normal to the layers, 
$\delta \vec{n} = - \vec{\nabla}_{\perp}\tilde{u}$
and an elastic energy $\frac{1}{2} K \vec{k}_{\perp} |\tilde{u}(\vec{k})|^{2}$.
Thus in smectic liquid crystals, the kernel
\begin{eqnarray}
v(\vec{k}) = B k_{||}^{2} + K \vec{k}_{\perp}^{4}.
\end{eqnarray} 
This is much alike our dispersion
for incommensurate phases apart
from a shift: in the incommensurate
phases the deviations were about
a finite wavenumber $\vec{q}$ whereas
in liquid crystal, the uniform $\vec{k}=0$
is the ground state.
As seen in the form
that we obtained for
$G(\vec{x})$,
the penetration depth $\lambda = \sqrt{K/B}$ determines the decay 
of an undulation distortion (splay director distortion)
imposed at the surface of the smectic. 

In conclusion, if $H_{soft}$ is indeed soft ($u \ll 1$)
then, in rotationally symmetric incommensurate soft spin XY systems,
quasi long range (algebraic) order may be observed at low temperatures,
These systems essentially may smectic thermodynamics.

\section{A Generalized
Mermin-Wagner-Coleman Theorem}
\label{M-W-low}

We will now  generalize the 
Mermin-Wagner-Coleman theorem 
\cite{Mermin-Wagner}, \cite{Coleman}:
{\bf All} continuous spin systems with translationally invariant 
two-spin interactions in two dimensions
with a rotationally
symmetric twice differentiable \cite{precision} Fourier transformed
kernel $v(\vec{k})$ show no spontaneous symmetry 
breaking at finite temperatures $T>0$. 
This is true for {\bf all} systems, irrespective of the 
range of the interaction or of its nature.
As will be made clear shortly, our analysis holds  
for both commensurate and incommensurate systems.

\subsection{The Classical Case}

Our approach is the 
standard one.  We will 
keep it more general
instead of specializing
to anti/ferromagnetic order 
or to interactions 
of one special sort. With the 
notations introduced 
in Section(\ref{definitions}),
we investigate $n$ component
spins on a lattice.
An applied magnetic field 
\begin{eqnarray}
\vec{h}(\vec{x}) = h \cos(\vec{q} \cdot \vec{x}) \hat{e}_{\alpha}
\end{eqnarray}
causes the spins
to take on their ground state values.

If $n=\alpha=2$ the unique spiral ground state ($\vec{S}^{g}(\vec{x})$), 
to which a low temperature  system would 
collapse to under the influence
of such a perturbation is
\begin{eqnarray}
S_{1}^{g}(\vec{x}) = \sin (\vec{q} \cdot \vec{x}).
\end{eqnarray}
When $n=3$ the ground state is not unique:
\begin{eqnarray}
S_{i<n}^{g}(\vec{x}) = r_{i} \sin(\vec{q} \cdot \vec{x}) \nonumber
\\ \sum_{i=1}^{n-1} r_{i}^{2}=1
\end{eqnarray}
and a magnetic field may be applied along two directions,
with all the ensuing steps trivially modified. 
With the magnetic field applied 
\begin{eqnarray}
H = \frac{1}{2} \sum_{\vec{x},\vec{y}} \sum_{i=1}^{n}
V(\vec{x}-\vec{y}) S_{i}(\vec{x}) S_{i}(\vec{y}) - 
\sum_{\vec{x}} h_{n}(\vec{x}) S_{n}(\vec{x}).
\label{last_H}
\end{eqnarray}
Note that the knowledge of the ground state
is not imperative in providing the forthcoming
proof \cite{smart} (whether a spiral or any
other state).  Our analysis thus holds  
for both commensurate and incommensurate systems.
We exploit the standard rotational
invariance of the measure
\begin{eqnarray}
\int d \mu~ ~\cdot    = ~ Z^{-1}  \int \prod_{\vec{x}} d^{n}S(\vec{x})
\delta(S^{2}(\vec{x})-1) e^{-\beta H}.
\end{eqnarray}
This is not applicable 
to many other spin models (e.g. 
Dzyaloshinskii-Moriya interactions, orbital Jahn-Teller
and Kugel-Khoskii models). The generators
 of rotation in the $[\alpha \beta]$ plane at a 
lattice site $\vec{x}$
are
\begin{eqnarray}
L^{\alpha \beta}_{\vec{x}} \equiv S_{\alpha}(\vec{x}) 
\frac{\partial}{\partial S^{\beta}(\vec{x})}
-S_{\beta}(\vec{x}) \frac{\partial}{\partial S^{\alpha}(\vec{x})}.
\label{Ld}
\end{eqnarray}
For any single spin, \begin{eqnarray}
0= \frac{d}{d \theta} \int d^{n}S ~\delta(\vec{S}^{2}-1)~
\nonumber
\\
f(S_{1},...,S_{\alpha} \cos \theta + S_{\beta} \sin \theta,...
\nonumber
\\, S_{\beta} \cos \theta - S_{\alpha} \sin \alpha,...,S_{n}).
\end{eqnarray}
It follows that 
\begin{eqnarray}
0= \int d^{n}S \delta (\vec{S}^{2}-1) L^{\alpha \beta} f(\vec{S}).
\label{L_ident}
\end{eqnarray}
In the up and coming, $\perp$ will denote the
the projection 
along the $\beta$ direction.
We define the operators
\begin{eqnarray}
\vec{A}(\vec{k}) \equiv \sum_{\vec{x}} \exp[i \vec{k} \cdot \vec{x}]
\vec{S}_{\perp}(\vec{x}), \nonumber
\\ \vec{B}(\vec{k}) \equiv \sum_{\vec{x}} \exp[i (\vec{k} + \vec{q}) 
\cdot \vec{x}]
\vec{L}_{\vec{x}}(\beta H),
\end{eqnarray}
where $\vec{L}_{x} \equiv \{ L^{i}_{\vec{x}} \}_{i=1}^{n-1}$ with 
$L^{i}_{\vec{x}} = L^{\alpha = n, \beta = i}_{\vec{x}}$.

By the Schwarz inequality,
\begin{eqnarray}
| \langle \sum_{i= \alpha, \beta} A_{i}^{*}B_{i} \rangle |^{2} ~\le ~~ \langle 
\sum_{i= \alpha, \beta} A_{i}^{*}A_{i} \rangle * 
\langle \sum_{i=\alpha,\beta} B_{i}^{*}B_{i} \rangle.
\end{eqnarray}
We will let $i=\alpha,\beta$ in the sum span a two element
subset of the $n$ spin components.
For any functional $C$:
\begin{eqnarray}
\vec{L}_{\vec{x}}(e^{-\beta H}C) = e^{-\beta H} \{ \vec{L}_{\vec{x}}(C)
+ C \vec{L}_{\vec{x}}(-\beta H)\}.
\end{eqnarray}
Invoking Eq.(\ref{L_ident}),
\begin{eqnarray}
0 = \int \prod_{\vec{x}}~ d^{n}S(\vec{x})~ \delta (\vec{S}^{2}(\vec{x})-1)
~\vec{L}_{\vec{x}}[e^{-\beta H}C].
\end{eqnarray}
Employing the last two equations in tow,
\begin{eqnarray}
\langle C B(\vec{p}) \rangle = \langle \sum_{\vec{x}} \exp[i \vec{p}
\cdot \vec{x}] ~\vec{L}_{\vec{x}}(C) \rangle.
\end{eqnarray}
It is readily seen that
\begin{eqnarray}
\sum_{i= \alpha,\beta}  \langle L_{\vec{x}}^{i}(L_{\vec{y}}^{i}(\beta H)) 
\rangle~ =
 \beta \langle [\sum_{i=\alpha,\beta} S_{i}(\vec{x}) S_{i}(\vec{y})
 V(\vec{x}-\vec{y})
\nonumber
\\ - h(\vec{x}) S_{n}(\vec{x})] \rangle,
\end{eqnarray}
and
\begin{eqnarray}
\langle \vec{B}(\vec{k})^{*} \cdot \vec{B}(\vec{k}) \rangle 
= \beta \sum_{\vec{x},\vec{y}} 
 \{ (\cos (\vec{k} + \vec{q}) \cdot (\vec{x}-\vec{y})-1) \nonumber
\\ \Big[ \sum_{i=\alpha, \beta} \langle S_{i}(\vec{x}) S_{i}(\vec{y})
\rangle  \Big] 
 V(\vec{x}-\vec{y}) \}   \nonumber
\\ -
h(\vec{x}) \langle S_{n}(\vec{x}) \rangle 
 ~~\ge~ 0
\end{eqnarray}

Henceforth, for simplicity, we specialize 
to $n=2$. Fourier expanding the interaction
kernel 
\begin{eqnarray}
V(\vec{x}-\vec{y}) = \frac{1}{N} \sum_{\vec{t}} v(\vec{t}) e^{i \vec{t} \cdot
(\vec{x}-\vec{y})},
\end{eqnarray} 
and substituting
\begin{eqnarray}
\langle \vec{S}(\vec{x}) \cdot \vec{S}(\vec{y}) \rangle = \frac{1}{N^{2}}
\sum_{\vec{u}} \langle |\vec{S}(\vec{u})|^{2} \rangle e^{i \vec{u} \cdot 
(\vec{x} - \vec{y}) },
\end{eqnarray}
we obtain 
\begin{eqnarray}
0 \le \langle \vec{B}(\vec{k})^{*} \cdot  \vec{B}(\vec{k}) \rangle
\equiv  
\beta \Delta_{\vec{k}}^{(2)} E -\beta h_{\vec{q}} \langle
S_{n}(-\vec{q}) \rangle = \nonumber
\\ \frac{\beta}{2N} \sum_{\vec{u}} 
 \Big[ v(\vec{u} + \vec{k}) + v(\vec{u}
- \vec{k}) \nonumber
\\ - 2 v(\vec{u}) \Big] \langle |\vec{S}(\vec{u})|^{2} \rangle
- \beta h_{\vec{q}} \langle S_{n}(-\vec{q}) \rangle 
\label{B*B}
\end{eqnarray}
where $\Delta_{\vec{k}}^{(2)} E$ measures 
the finite difference 
of the internal energy with respect to 
a boost of momentum $\vec{k}$. 
\begin{eqnarray}
\langle \vec{A}(\vec{k})^{*} \cdot \vec{A}(\vec{k}) \rangle = 
\sum_{\vec{x},\vec{y}}
\langle \vec{S}_{\perp}(\vec{x}) \cdot \vec{S}_{\perp}(\vec{y})
\rangle
\nonumber
\\ \times \exp[i \vec{k} \cdot (\vec{x}-\vec{y})].
\end{eqnarray}
\begin{eqnarray}
 \langle \vec{A}(\vec{k})^{*} \cdot \vec{B}(\vec{k}) \rangle= 
\langle \sum_{i,\vec{x}}
L_{\vec{x}}^{i}(\vec{S}_{\perp}(\vec{x})) \exp[i (\vec{k} + \vec{q}) 
\cdot \vec{x}] \rangle
= m_{\vec{q}},
\nonumber
\end{eqnarray}
where
$m_{\vec{q}} \equiv \langle S_{n}(\vec{q}) \rangle$
and, as noted earlier, $\perp$ refers to the $i=1$ spin direction 
orthogonal to $i=n=2$.
Note that with our convention
for the Fourier transformations,
a macroscopically modulated state of
wave-vector $\vec{q}$, the magnetization
$m_{q} ={\cal{O}}(N)$
as is the energy difference in Eq.(\ref{B*B}).
Trivially rewriting the 
Schwarz inequality and
summing over all
momenta $\vec{k}$,
\begin{eqnarray}
\sum_{\vec{k}} \frac{
\langle \vec{A}(\vec{k})^{*} \cdot \vec{B}(\vec{k})
\rangle}{\langle 
\vec{B}(\vec{k})^{*} \cdot \vec{B}(\vec{k}) \rangle}
\le \sum_{\vec{k}} \langle \vec{A}(\vec{k})^{*} \cdot \vec{A}(\vec{k})
\end{eqnarray}
which explicitly reads
\begin{eqnarray}
2 N |m_{\vec{q}}|^{2}  \Big( \beta  \sum_{\vec{k}} 
 ( \langle |\vec{S}(\vec{u})|^{2} \rangle  [v(\vec{k}+\vec{u}) \nonumber 
\\ +v(\vec{u}- \vec{k})-2 v(\vec{u})]+ 2 |h||m_{q}| 
\Big)^{-1}\nonumber
\\ \le N \sum_{\vec{k}} \sum_{\vec{x},\vec{y}} \langle \vec{S}_{\perp}(\vec{x})
\cdot \vec{S}_{\perp}(\vec{y}) \rangle e^{i \vec{k} \cdot (\vec{x}-\vec{y})}
\nonumber 
\\ = N \sum_{\vec{x}} \langle \vec{S}_{\perp}^{2}(\vec{x}) \rangle.
\label{ineq}
\end{eqnarray}
Explicitly, as the integral $\int^{|\vec{k}| > \delta}$ 
$\frac{d^{d}k}{(2 \pi)^{d}} ...$ is non-negative 
(as $\langle \vec{B}(\vec{k})^{*} \cdot \vec{B}(\vec{k}) \rangle \ge 0$
the denominator in Eq.(\ref{ineq})
is positive for each individual 
value of $\vec{k}$), and
as $\langle \vec{S}_{\perp}^{2}(\vec{x}) \rangle  \le 1$,
we obtain in the thermodynamic limit
\begin{eqnarray}
\frac{2}{\beta} |m_{\vec{q}}|^{2}  \int^{|\vec{k}| < \delta}
\frac{d^{d}k}{(2 \pi)^{d}} \Big[ \int \frac{d^{d}u}{(2 \pi)^{d}} \nonumber
\\  \langle |\vec{S}(\vec{u})|^{2} \rangle  (v(\vec{k}+\vec{u}) 
+v(\vec{k}-\vec{u})-2 v(\vec{u}))  \nonumber
\\ +2 |h| |m_{q}| \Big]^{-1}
 \le 1.
\label{main}
\end{eqnarray}
Taking $\delta$ to be small we may bound from above 
(for each value of $\vec{k}$) the positive
denominator in the square brackets 
and consequently
\begin{eqnarray}
 \int^{|\vec{k}| < \delta}
\frac{d^{d}k}{(2 \pi)^{d}} \Big[ \int \frac{d^{d}u}{(2 \pi)^{d}} \nonumber
\\  \langle |\vec{S}(\vec{u})|^{2} \rangle  (v(\vec{k}+\vec{u}) 
+v(\vec{k}-\vec{u})-2 v(\vec{u}))  \nonumber
\\ +2 |h| |m_{q}| \Big]^{-1}
\ge \int^{|\vec{k}| < \delta}
\frac{d^{d}k}{(2 \pi)^{d}} \Big[ \int \frac{d^{d}u}{(2 \pi)^{d}} \nonumber
\\ A_{1}  k^{2} \lambda_{\vec{u}} \langle |\vec{S}(\vec{u})|^{2} \rangle  
+2 |h| |m_{q}| \Big]^{-1},
\end{eqnarray}
with $\lambda_{\vec{u}}$ chosen to be the largest 
principal eigenvalue of the $d \times d$ 
matrix $\partial_{i} \partial_{j} [v(\vec{u})]$,
and $A_{1}$ a constant. 
For a twice differentiable $v(\vec{u})$, and for $|\vec{k}| \le \delta$ 
where $\delta$ is finite,
\begin{eqnarray}
(v(\vec{k}+\vec{u}) 
+v(\vec{k}-\vec{u})-2 v(\vec{u})) 
\le A_{1} \lambda_{\vec{u}} k^{2} \le B_{1} k^{2}
\end{eqnarray}
for all $\vec{u}$ within the Brillouin Zone  
with the additional positive constant $B_{1}$
introduced \cite{precision}. Here we
reiterate that $\langle \vec{B}^{*}(\vec{k}) \cdot \vec{B}(\vec{k}) \rangle$
of Eq.(\ref{B*B}) is positive definite
and consequently the
bound derived is 
powerful.

In $d \le 2$, the integral
$\int^{|\vec{k}| < \delta} \frac{d^{d}k}{B_{1}k^{2}}$
diverges making it possible to satisfy  
Eq.(\ref{main}), at finite temperatures,
when the external magnetic
field $h \rightarrow 0$ only if 
the magnetization $m_{q}=0$.
If finite size effects 
are restored, in a system
of size $N= L \times L$ where 
the infrared cutoff in
the integral is ${\cal{O}} (\frac{2 \pi}{L}, \frac{2 \pi}{L})$
the latter integral diverges as ${\cal{O}}(\ln N)$.
This implies that the upper bound on $|m_{\vec{q}}|$
scales as ${\cal{O}}(N / \sqrt{\ln N})$, much lower than
the $ {\cal{O}}(N)$ requisite for finite 
on-site magnetization. For further 
details see \cite{smart}. 
If there are $M \ge 2$ pairs
of minimizing modes and $2p +1 \ge  n \ge 2p$
(with an integer $p$) then we may apply an infinitesimal 
symmetry breaking magnetic field along, at most, $\min\{p,M\}$ 
independent spin directions ($\alpha$). Employing the
spin rotational invariance within each  plane 
 $[\alpha \beta]$ associated with any individual 
mode $\vec{\ell}$ (au lieu of a specific $\vec{q}$)
we may produce a bound similar that 
in Eq.(\ref{main}) wherein $\langle |\vec{S}(\vec{u})|^{2}
\rangle$ will be replaced by $\sum_{i= \alpha,\beta} 
\langle |S_{i}(\vec{u})|^{2} \rangle$ and $|m_{\vec{q}}| \rightarrow
|m_{\alpha}(\vec{\ell})|$.

\subsection{The Quantum Case}

The finite temperature behavior 
of a quantum system is, in many 
respects, similar to that of a 
classical system. The quantum system
is also invariant under rotations 
with $\hat{S}^{2}(\vec{x}) = S(S+1)$.
Alternatively, one could directly tackle the
$n=3$ quantum case by applying the Bogoliubov
inequality
\begin{eqnarray}
\frac{\beta}{2} \langle \{A,A^{\dagger}\} \rangle * \langle
[~[C,H],C^{\dagger}]\rangle \ge | \langle [C, A]\rangle |^{2}
\end{eqnarray}

with $[~,~]$ and $\{~,~\}$ the commutator and anticommutator
respectively. From this inequality,
it follows that $ \langle
[~[C,H],C^{\dagger}]\rangle$
is positive definite. 
In particular, for any six
operators $\{A_{1},A_{2},A_{3},C_{1},C_{2},C_{3}\}$,
\begin{eqnarray}
\frac{\beta}{2} (\sum_{a=1}^{3} 
 \langle \{A_{a},A_{a}^{\dagger}\} \rangle)
(\sum_{a} \langle
[~[C_{a},H],C^{\dagger}_{a}]\rangle)
\ge  \nonumber
\\ \sum_{a} | \langle [C_{i}, A_{i}]\rangle |^{2}
\label{Bog_symm}
\end{eqnarray}

Setting $A_{1}= S_{2}(\vec{q}-\vec{k})$ and 
$C_{1}= S_{1}(\vec{k})$
we will once
again obtain Eq(\ref{main}) 
with the classical spins replaced 
by their quantum counterparts.

Rather explicitly, employing  
\begin{eqnarray}
[S^{\alpha}(\vec{k}), S^{\beta}(\vec{k}^{\prime})] = i 
\epsilon^{\alpha \beta \gamma} S_{\gamma}(\vec{k}+ \vec{k}^{\prime}),
\end{eqnarray}
we find for the Hamiltonian
of Eq.(\ref{last_H}) (with $n=3$),
\begin{eqnarray}
[[C_{1},H],C_{1}^{\dagger}] = 
\frac{1}{2N} \sum_{\vec{k}^{\prime}} 
(S_{2}(\vec{k}^{\prime}) S_{2}(-\vec{k}^{\prime})
+ S_{3}(\vec{k}^{\prime}) S_{3}(-\vec{k}^{\prime})) \nonumber
\\ \times
[v(\vec{k}+ \vec{k}^{\prime}) - 2 v(\vec{k}) 
+ v(\vec{k}^{\prime} - \vec{k})] \nonumber
\\ 
+ \frac{1}{N} \sum_{\vec{k}^{\prime}} h_{3}(\vec{k}^{\prime}) 
S_{3}(-\vec{k}^{\prime}). 
\nonumber
\end{eqnarray}

Similarly,
\begin{eqnarray}
|\langle [C_{1}, A_{1} ] \rangle|^{2} = |\langle S_{3}(\vec{q}) \rangle|^{2},
\end{eqnarray}
the squared magnetization along the z (or 3) direction for a
mode $\vec{q}$, 
and
\begin{eqnarray}
\sum_{\vec{k}} \langle \{ A_{1}, A^{\dagger}_{1} \} \rangle
= 2 \sum_{\vec{k}} \langle S_{2}(\vec{k}) S_{2}(-\vec{k}) \rangle \nonumber
\\  = 2 \langle [S_{2}(\vec{x}=0)]^{2} \rangle.
\end{eqnarray}
Next, let us cyclically
set, $A_{2} = S_{3}(\vec{q}-\vec{k}), 
C_{2} = S_{2}(\vec{k}), A_{3} = S_{1}(\vec{q} - \vec{k})$, 
and $C_{3} = S_{3}(\vec{k})$.
The commutators associated
with these operators
are all identically the same
apart from a uniform 
cyclic permutation
of all spin components
involved.
Trivially rewriting 
the symmetrized Bogoliubov inequality 
Eq.(\ref{Bog_symm})
and summing over all
modes $\vec{k}$,
\begin{eqnarray}
\frac{\beta}{2} \sum_{\vec{k},a} 
\langle \{A_{a}, A^{\dagger}_{a} \} \rangle
\ge \sum_{\vec{k}} \frac{  \sum_{a}| \langle  
[C_{a}, A_{a}] \rangle |^{2}}{ \sum_{a} \langle [[C_{a},H],
C^{\dagger}_{a}] \rangle}.
\end{eqnarray}
Replacing the $\vec{k}$ sums by
integrals in the thermodynamic
limit, and
employing 
the positivity
of $\langle [[C_{a},H],
C_{a}^{\dagger}] \rangle$
that follows from the
Bogoliubov inequality
for each individual
value of $\vec{k}$, we find the 
quantum analogue of
Eq.(\ref{main}),
\begin{eqnarray}
\frac{1}{2 \beta} |m_{\vec{q}}|^{2}  \int^{|\vec{k}| < \delta}
\frac{d^{d}k}{(2 \pi)^{d}} \Big[ \int \frac{d^{d}u}{(2 \pi)^{d}} \nonumber
\\  \langle |\vec{S}(\vec{u})|^{2} \rangle  (v(\vec{k}+\vec{u}) 
+v(\vec{k}-\vec{u})-2 v(\vec{u}))  \nonumber
\\ +|h| |m_{q}| \Big]^{-1}
 \le S(S+1).
\label{main_q}
\end{eqnarray}
Apart from the simple scaling
factor of $4S(S+1) = 2 \sum_{a} 
\langle \{A_{a}, A^{\dagger}_{a} \} \rangle$
by comparison to the classical case,
there is no difference
between this inequality
and its classical counterpart
in the zero field limit.
We symmetrized the Bogoliubov inequality
(Eq.(\ref{Bog_symm})) in order
to avoid the appearance
of only two transverse 
spin components in the
Fourier weights $\langle |S_{i}(\vec{q})|^{2} 
\rangle$ so as to give 
the resulting Mermin-Wagner inequality
a transparent physical meaning
associated with the energy 
boost differences $\Delta_{\vec{k}}^{(2)} E$
which are symmetric in all
spin indices. From here onward the discussion 
can proceed as in the classical
case.

\section{Mermin-Wagner Bounds in High Dimensions}
\label{M-W-I-boost}

In any dimension,  
Eq.(\ref{main}) reads in the limit $h \to  0$ 
\begin{eqnarray}
2 |m_{\vec{q}}|^{2} T \int \frac{d^{d}k}{(2 \pi)^{d}} ~ ~ 
\frac{1}{\Delta_{\vec{k}}^{(2)}E} \le 1,
\label{general**}
\end{eqnarray}
with the shorthand defined by
Eq.(\ref{B*B}).
By parity invariance and noting that   
$\{\vec{S}(\vec{x})\}$ are real,
\begin{eqnarray}
\sum_{\vec{u}} v(\vec{k} + \vec{u}) \langle
|\vec{S}(\vec{u})|^{2} \rangle = \sum_{\vec{u}} v(\vec{k} - \vec{u})
\langle
|\vec{S}(\vec{u})|^{2} \rangle.
\end{eqnarray}
Thus the denominator of Eq.(\ref{general**}) 
reads $\Delta_{\vec{k}}^{(2)}E = [2(E_{\vec{k}}-E_{0})]$
where $E_{\vec{k}}$ is the internal 
energy of system after undergoing a 
boost of momentum $\vec{k}$ and $E_{0}$ 
denotes the internal energy of the 
un-boosted system.
In some instances when the dispersion relation 
about an assumed zero temperature ground
state is inserted into Eq.(\ref{general**}),
we will find that the integral in Eq.(\ref{general**})
diverges: At arbitrarily
low temperatures we cannot assume the zero temperature
ground state with the natural dispersion relation
$\Delta E$ for fluctuations about it.
A case in point is the dispersion relation
for the Coulomb
Frustrated Ferromagnet. The denominator
in Eq.(\ref{main}) is a finite temperature 
extension of the $T=0$ dispersion  
relations ($\Delta_{\vec{k}}^{(2)} E$).
In general, at zero temperature,
\begin{eqnarray}
\langle |\vec{S}(\vec{k})|^{2} \rangle = \frac{N^{2}}{2}
[\delta_{\vec{k},\vec{q}}+ \delta_{\vec{k},-\vec{q}}]
\label{ideal-occ}
\end{eqnarray}
and the integral 
in Eq.(\ref{general**})
becomes 
\begin{eqnarray}
\int \frac{d^{d}k}{(2 \pi)^{d}} \frac{1}{v(\vec{k}+\vec{q}) 
+ v(\vec{k}-\vec{q}) - 2 v(\vec{q})}
\label{MY_IN}.
\end{eqnarray}
For the incommensurate models,
the reader will recognize
this as none other
than the integral
obtained by other methods 
via the Dirac like analysis
in Section(\ref{XY-fluct}).
Specifically, the denominator
of Eq.(\ref{MY_IN}) is
$E_{low}$ of Eq.(\ref{spec})
wherein $\vec{k}_{1} = \vec{q} + \vec{k}$
and $\vec{k}_{2} = - \vec{q} + \vec{k}$
(which transformed into 
Eq.(\ref{twist*})  (with $\vec{\delta}$ 
now portrayed by $\vec{k}$) for general
rotationally symmetric
incommensurate systems).

Whenever Eq.(\ref{MY_IN}) diverges, an assumption
of an almost ordered ground state at arbitrarily
low temperatures ($T = 0^{+}$) 
with the ensuing zero temperature 
dispersion relation 
($\Delta_{\vec{k}}^{(2)}E = [2(E_{\vec{k}}-E_{0})]$) 
about that ordered state is flawed: Eq.(\ref{general**}) 
is strongly violated.
This poses no problem for most 
canonical $ d>2$ dimensional models
where the integral in Eq.(\ref{MY_IN})
is finite. For all three dimensional incommensurate systems
with a rotational symmetry that secures, as $q \neq 0$, 
a shell of radius q of minimizing modes about 
the origin (e.g. the  three dimensional 
Coulomb Frustrated Ferromagnet and other 
high dimensional systems),
the divergence of this integral
hints possible non-trivialities. 
The decoherence time 
scale (or more precisely,
bandwidth time scale \cite{moessner}) $\langle \tau \rangle $ 
may diverge if we assume a 
zero temperature 
dispersion of fluctuations
about an ideal crystal.
A divergent decoherence time scale suggests the 
absence of broken translational
symmetry. Formally,
\begin{eqnarray}
\int \frac{d^{3}k}{(2 \pi)^{3}} \tau_{k} = \tau(\vec{r}=0).
\end{eqnarray} 
In order
for magnetization $m_{q} = {\cal{O}}(N)$
to arise, the average of the inverse boost energy
over all of $\vec{k}$ space
\begin{eqnarray}
\tau \equiv \langle \frac{1}{\Delta_{\vec{k}}^{(2)}E} 
\rangle_{\vec{k}} \le  {\cal{O}}(T^{-1}),
\label{last-ate}
\end{eqnarray}
at all $0<T<T_{c}$.
In most $d >2$ systems
this is trivially
satisfied with the average
bounded by a constant at zero temperature.
This average 
diverges at $T=0$ whenever the integral
of Eq.(\ref{MY_IN}) does.
There are two possibilities:

(i) The system is ordered at all temperatures $T<T_{c}$
in which case, the thermodynamic average of Eq.(\ref{last-ate}) 
is finite for all $T>0$ and $\tau$ is non-analytic 
at $T=0$.

(ii) The system is disordered at all finite temperatures
and orders classically only at $T=0$.

The first possibility ((i)) was argued for
by a non-rigorous yet  
elegant diagrammatic analysis
by Brazovskii \cite{Brazovskii} 
long ago: Thermal fluctuations,
on their own, may enhance (or generate) cubic terms fortifying
(or triggering weak) first order transitions.
This cannot be ruled out by
the rigorous Mermin-Wagner inequalities
that we derived.
As reiterated, if the fluctuation integral
diverges then we may not obtain the 
low temperature $T=0^{+}$ 
dispersion by assuming a 
nearly perfectly ordered 
state. This does not
rigorously preclude order.
Order, if it exists at
arbitrarily low temperatures
must display non-trivial 
excitation spectra about it
having a vital explicit thermal
dependence. 
If such a 
possibility arises,
it might be immaterial
if the system is permanently
frozen into a glass before
reaching an equilibrium 
thermodynamic transition \cite{new}.
In both cases, $\tau(T)$ is
non-analytic at $T=0$.
The integral of Eqs.(\ref{MY_IN},\ref{last-ate}) 
has a suggestive physical interpretation. If 
the quantum spin system is subjected to a boost of momentum $\vec{k}$, then 
\begin{eqnarray}
1/\Delta_{\vec{k}}^{(2)}E \equiv \tau_{\vec{k}}
\end{eqnarray}
is the characteristic lifetime of the
excited state. The average in Eqs.(\ref{last-ate}) is 
the characteristic relaxation (or decoherence) time 
of the system
averaged over magnons of all possible 
momenta. Whenever the average characteristic relaxation
time 
\begin{eqnarray}
\langle \tau_{\vec{k}} \rangle = \int
\frac{d^{d}k}{(2 \pi)^{d}} \Big[ \int \frac{d^{d}u}{(2 \pi)^{d}} \nonumber
\\  \langle |\vec{S}(\vec{u})|^{2} \rangle  (v(\vec{k}+\vec{u}) 
+v(\vec{k}-\vec{u})-2 v(\vec{u})) \Big]^{-1}
\end{eqnarray}
diverges then by our generalized 
inequality 
\begin{eqnarray}
2|m_{q}|^{2}T \langle \tau_{\vec{k}} \rangle = 2|m_{q}|^{2}T  
\tau_{r=0} \le 1
\end{eqnarray}
the system does not order
in such a way that the 
fluctuation dispersion
about any viable ground
state is valid. 
We emphasize that, even
in case (ii), the Mermin-Wagner 
inequality allows for 
algebraic long range
order. Indeed, we have
found by our primitive $n=2$ analysis
(assuming a fluctuation
spectrum about a $T=0$ ordered
state), that although the average
magnetization vanishes at all $T >0$,
the correlations decay algebraically
(see Eq.(\ref{ALRO}) and the appendix
where this correlator is explicitly 
evaluated). As fortold, the main difference 
between the results of our
rigorous analysis here
and that of the soft 
spin model in section(\ref{XY-fluct})
is that the soft spin analysis
relied on a comparison
to the ground state
so that no explicit thermal 
dependence of the mode occupancies
$\langle |\vec{S}(\vec{k})|^{2} \rangle$
was allowed. As fortold,
if the mode occupancies have
a non-trivial thermal 
dependence which cannot be
neglected even at arbitrarily
low $T=0^{+}$ temperatures, ordering 
may occur.

\section{Classical $O(n=3)$ fluctuations}
\label{n=3}

We now return to the more naive 
``soft-spin''  (see section (\ref{XY-fluct}) for the soft spin XY)
$O(3)$ models in order to witness an intriguing 
even-odd binding-unbinding 
(or pairing - unpairing) effect 
that could have otherwise been 
missed.  We will see that
$H_{1}$  plays the role 
of a pairing interaction.
We will see the incommensurate classical even n soft-spin systems
are more "gapped" than their odd n counterparts. When, in three 
dimensions, incommensurate continuous spin systems are endowed with a 
spatial rotational symmetry which secures a finite shell of minimizing modes, 
odd $n$ systems may be disordered while their even $n$ counterparts
may exhibit, by the same soft-spin analysis, critical algebraic 
long range order. This even/odd alteration is reminiscent of the 
appearance/non-appearance of a gap for quantum spin $s$ chains 
with even/odd values of $(2s+1)$ 
\cite{Haldane}, \cite{Fradkin} and the physics for an 
even/odd number of legs in 
an $s=1/2$ quantum spin ladder \cite{sudip} where
such even/odd discrepancies are triggered by
Berry phase terms. As before, we will begin
our analysis for general incommensurate
soft spin models and only at
the end investigate what occurs
for rotationally symmetric
systems with a shell of 
minimizing modes.
As we proved, for an $n=3$ system the generic 
ground states are simple spirals.
If we rotate the helical ground-state 
to the $1-2$ plane; the single quadratic 
term in $\delta S_{i=3}(\vec{x})$ 
is $\sum_{\vec{x}} (\delta S_{i=3}(\vec{x}))^{2}$.
The eigenvalues $\lambda_{i \ge 3} = 2 =  \lambda_{-} = \lambda_{\min}$
(see Eq.(\ref{sep}). Here, all that follows holds for
arbitrarily large $u$- the only approximation that 
we are making is neglecting $O((\delta S)^{3})$
terms by comparison to quadratic terms: i.e. assuming
that $\delta S(\vec{x}) \ll 1$. 
Unlike the above treatment of the XY spins,
no small $u$ is necessary in order to 
make headway on the Heisenberg problem. 
For $n=3$ we find that the fluctuation
eigenstates of the are the products 
of an eigenstate of ${\cal{H}}$ within the plane 
of the spiral ground state  
and a fluctuation  eigenstate of $v(\vec{k})$
along the direction orthogonal
to the spiral plane. Written formally, 
to quadratic order, the fluctuation eigenstates are
$|\psi_{m} \rangle  \otimes |\delta S_{3}(\vec{k}) \rangle$.
Fluctuations along the $i=3$ axis are orthogonal (in a geometrical and
formal sense) to the ground-state plane. This is expected 
as fluctuations in any hyper-plane perpendicular to the $[12]$ plane
do not change, to lowest order, the norm of the spin. 
$|\delta S_{3}(\vec{k}) \rangle$ is literally the ``odd'' man out.
As foretold, this is a general occurrence. Whenever the number
of spin  components is odd, one unpaired spin component 
is unaffected by the interaction enforcing 
the spin normalization constraint. 

Within the $i=3$ subspace $\langle \vec{k}| M |\vec{k}^{\prime} \rangle =
2 \delta_{\vec{k},\vec{k}^{\prime}}$ ~and our previous analysis follows.
The dispersion $E_{k}^{i=3} = [v(\vec{k})+ 2~ u ~ \lambda_{i=3}]$
does not have a higher minimum than with $E_{m}$ \cite{explain}. 
Both $E_{k}^{i=3}$ and $E_{m}$ share the same value of $\lambda$
and consequently the $\delta S_{i=3}(\vec{x})$
fluctuations are minimized at wave-vectors $ \vec{\ell} \in M_{Q}$ such that
$v(\vec{\ell}) = v(\vec{q}) = \min_{\vec{k}} v(\vec{k})$. 
As $|\psi_{m} \rangle$ is a normalized superposition
of $|\delta \vec{S}(\vec{k}) \rangle$ modes
(the latter spin vectors being in the $1-2$ plane) 
and $v(\vec{k})$ is diagonal, 
and attains its minimum 
at $\vec{\ell} \in M_{Q}$:
\begin{equation}
\min_{m} \{ E_{m} \} \ge E_{\vec{\ell}\in M_{Q}}^{i=3}.
\end{equation}
Here, we invoked the trivial inequality 
\begin{eqnarray}
\min_{\psi} \langle \psi| [H_{0}+H_{1}]| \psi \rangle  \ge \min_{\phi}
\langle \phi|H_{0}|\phi \rangle + 
\min_{\xi}  \langle \xi|H_{1}|\xi \rangle. 
\end{eqnarray} 
Thus, there exist Goldstone modes corresponding to $\delta S_{i=3}$ 
fluctuations, and one
must adjust additive constants such that
$\min_{\vec{k}}\{ E_{\vec{k}}^{i=3} \}= 0$.
The resulting fluctuation integral reads
\begin{equation}
\langle [\delta S_{i=3}(\vec{x}=0)]^{2} \rangle = 
k_{B}T \int \frac{d^{3}k}{(2 \pi)^{3}}~~ \frac{1}{v(\vec{k})
-v(\vec{q})}, 
\label{fluctuations}
\end{equation}
where we invoke $\langle \Delta S_{i}(\vec{k})
  \Delta S_{i}(\vec{k}^{\prime}) \rangle =
  \delta_{\vec{k}+\vec{k}^{\prime},0} \langle |\Delta
S_i(\vec{k})|^{2} \rangle$
by translational invariance,
and employed equipartition. Till now we investigated
general incommensurate systems,
we now specialize to rotationally
symmetric systems. As pointed out earlier,
when the incommensurate
spin system possesses 
a rotational symmetry,
the minimizing manifold $M$
is a $(d-1)$ dimensional
shell of radius $q>0$. The fluctuation integral receives divergent
contributions from the low energy modes nearby. 
By quadratic expansion about the minimum along $\hat{n} \perp M$,
a divergent one-dimensional integral for the bounded
 $\langle (\Delta \vec{S}(\vec{x}=0))^{2} \rangle$ signals that
are quadratic fluctuation analysis calculation is
inconsistent. We are led to the conclusion that
higher order constraining terms are imperative:
We cannot throw away cubic and quartic spin fluctuation terms
$(\delta S^{3,4}(\vec{x}))$
relative to the quadratic  $(\delta S^{2}(\vec{x}))$
terms (notwithstanding the fact
that all of these terms appear with $O(u)$ prefactors 
irrespective of how large $u$ is). The spin fluctuations 
$\delta S_{i=3}(\vec{x})$ are of order unity at all finite
temperatures and  {\bf{$T_{c}(q>0) \simeq 0$}}.
Note that this is ``almost  a theorem''. 
Here we do not demand that $u$ be small
(only $\delta S$). This is an important point.
$|\psi_{m} \rangle$ is an eigenstate for
arbitrarily large $u$.  To quadratic order
in the fluctuations,
the minimum belongs to 
$|\delta S_{3} \rangle $
(or is degenerate with it). 
The divergent fluctuations here
signal that ${\cal{O}}(\delta S^{4}) ={\cal{O}}(\delta S^{2})$.
Thus assuming that $\delta S \lesssim \sqrt{J/u}$
(with $J=1$ the exchange constant) we  reach a 
paradox. Thus, if the integral in Eq.(\ref{fluctuations})
diverges for an $O(3)$ model then  at all finite $T$:
$\delta S \ge \sqrt{J/u}$. Unlike the case
for $n=2$, there is no algebraic long range order.
Computing the correlator, we find that $G(\vec{x})$ 
decays exponentially at all temperatures (much
unlike the algebraic long range order
seen in Eq.(\ref{ALRO})).

When the incommensurability $q$ vanishes, the minimizing manifold 
shrinks to a point- the number of
nearby low energy modes is small 
and our fluctuation integral converges
in $d>2$. This is in accord with the
well known finite temperature phase 
transition of the nearest neighbor
Heisenberg ferromagnet: $T_{c}(q=0) = O(1)$
(or when dimensions are fully restored-
it is of the order of the exchange constant).
In many rotationally symmetric incommensurate
spin systems (such as the Coulomb Frustrated Ferromagnet
parameterized by the Coulomb strength $Q=q^{4}$), 
a discontinuity in $T_{c}$ may occur as
$M_{Q} \rightarrow M_{Q=0} \equiv(\vec{q} = 0)$.
We anticipate that small lattice corrections 
($\lambda \neq 0$) in $v_{Q}(\vec{k})$ to 
yield insignificant modifications to $T_{c}(Q)$: 
one way to intuit this is to estimate $T_{c}$ by
the temperature at which the fluctuations,
as computed within the quadratic Hamiltonian  
 $\langle \delta \vec{S}^{2}(\vec{x}=0) \rangle = O(1)$.

To summarize, we argued that in (essentially
hard spin) Heisenberg realizations
of sufficiently frustrated incommensurate
models, no conventional long range order is possible 
in the continuum limit at finite temperatures. 
More specifically, it is suggested that
if the integral $\int \frac{d^{d}k}{v(\vec{k})-v(\vec{q})}$
diverges then no long range order is possible at
finite temperature, unless an expansion
about the ground state is void (the thermal
dependence of the higher energy mode occupancies
cannot be neglected even at $T=0^{+}$) . 
If lattice effects are mild then 
$T_{c}(q>0)$ is expected to be small.
In the unfrustrated ($q=0$) Heisenberg
ferromagnet in $d=3$: $T_{c} = O(1)$.
Fusing these facts, a discontinuity in $T_{c}$
in all rotationally symmetric
incommensurate spin systems:
\begin{eqnarray}
 \delta T_{c} \equiv \lim_{q \rightarrow 0}[T_{c}(0) - T_{c}(q)]
\end{eqnarray}
is suggested by soft-spin analysis.
Such a discontinuity will occur as $Q \to 0$
in the Coulomb frustrated ferromagnet.

\section{$O(n \ge 4)$ fluctuations}
\label{n=4}

We will now find that, on the level of lowest order
spin wave fluctuation analysis, algebraic long range order
might be possible for $n=4$ component spins
even when their $n=3$ counterparts may exhibit exponential
decay of correlations at all finite temperatures
for the same Hamiltonian.
The fluctuation analysis of any $O(n>2)$ system
about a spiral ground state is qualitatively similar
to that of the Heisenberg system.
As noted previously, poly-spiral states
will tend to dominate at large $n$.
The reader can convince him/herself that 
for even $n$ with a $p<n/2$ poly-spiral ground state
and for  all odd $n$  the fluctuations will give 
rise to a leading order $\epsilon^{-1}$ divergence. 
The reasoning is simple: the poly-spiral states extend
along an even number of axis. If $n$
is odd then there will be at least
one internal spin direction $i$ along 
which $S_{i}^{g}=0$  and our analysis of the
Heisenberg model can be 
reproduced. The lowest 
eigen-energy associated
with the fluctuations $|\psi_{m} \rangle$
in the $(2p)$ dimensional
space spanned by the ground state
is higher than 
the lowest eigen-energy 
for fluctuations along 
an orthogonal direction. The term of constraint is positive definite,
$\langle \psi|H_{soft}|\psi \rangle \ge 0$.  If $|\delta S| \ll 1$,
this implies that the quadratic term in $\delta S(\vec{x})$ 
stemming from $H_{1}$ is non-negative definite.
For $S_{i}^{g}(\vec{x}) =0$, this quadratic term in 
$\delta S_{i}(\vec{x})$
is zero. The eigenvalue $\lambda_{\min} = \lambda_{-} =2$ 
corresponds to the zero contribution in 
$O(\delta S^{2}(\vec{x}))$ from $H_{1}$.  
Once again, $\min_{m} \{ E_{m} \} \ge E_{\vec{\ell}\in M_{Q}}^{i}$,
from the simple
\begin{eqnarray}
\min_{\psi} \langle \psi| [H_{0}+H_{soft}]| \psi \rangle  \ge 
\nonumber
\\ \min_{\phi} \langle \phi|H_{0}|\phi \rangle  + 
\min_{\xi} \langle \xi|H_{soft}|\xi \rangle. 
\end{eqnarray} 
The fluctuations of even component
spin about a $p=n/2$ poly-spiral 
ground state are more complicated.
As before, coupling between different
modes occurs. In this case they are more 
numerous. For $n=4$, the fluctuation energy, to quadratic order,
about a bi-spiral reads 

$
\! \! \! \! \! \! \! \! \! \delta H = 
\frac{1}{2N} \sum_{\vec{k}} [v(\vec{k}) - v(\vec{q})] 
|\delta \vec{S}(\vec{k})|^{2} 
+ \frac{u a_{1} a_{2}}{N} \{ \sum_{\vec{k}_{1},\vec{k}_{2}}
[ (\delta S_{1}(\vec{k}_{1}) \delta S_{3} (\vec{k}_{2}) 
+ \delta S_{1}(\vec{k}_{2})
\delta S_{3}(\vec{k}_{1})) \nonumber
\\ \times 
[\delta _{\vec{q}_{1}+\vec{q}_{2}+\vec{k}_{1}+\vec{k}_{2},0} 
+ \delta_{\vec{q}_{1}+\vec{q}_{2}-\vec{k}_{1}-\vec{k}_{2},0}
 + \delta_{\vec{k}_{1}+\vec{k}_{2}+\vec{q}_{2}-\vec{q}_{1},0} +
\delta_{\vec{k}_{1} +\vec{k}_{2} +\vec{q}_{1}-\vec{q}_{2},0}] +  \nonumber
\\ + (\delta S_{2}(\vec{k}_{2}) \delta S_{4}(\vec{k}_{1})+
\delta S_{2}(\vec{k})\delta S_{4}(\vec{k}_{2}))\nonumber
\\ \times [\delta_{\vec{k}_{1}+\vec{k}_{2}+\vec{q}_{1}-\vec{q}_{2},0}+
\delta_{\vec{k}_{1}
+\vec{k}_{2}+\vec{q}_{2}-\vec{q}_{1},0}-
\delta_{\vec{k}_{1}+\vec{k}_{2}+\vec{q}_{1}+
\vec{q}_{2},0}-\delta_{\vec{k}_{1}+\vec{k}_{2}-
\vec{q}_{1}-\vec{q}_{2},0}] \nonumber
\\ -i (\delta S_{1}(\vec{k}_{2}) \delta S_{4}(\vec{k}_{1})
+ \delta S_{1}(\vec{k}_{1}) \delta S_{4}(\vec{k}_{2})) \nonumber
\\ \times  [\delta_{\vec{k}_{1}+\vec{k}_{2}+\vec{q}_{1}+\vec{q}_{2},0}-
\delta_{\vec{k}_{1}+\vec{k}_{2}-\vec{q}_{1}-\vec{q}_{2},0} +
\delta_{\vec{k}_{1}+\vec{k}_{2}+\vec{q}_{2}-\vec{q}_{1},0}-
\delta_{\vec{k}_{1}+\vec{k}_{2}-\vec{q}_{2}+\vec{q}_{1},0}] \nonumber
\\ -i (\delta S_{2}(\vec{k}_{2}) \delta S_{3}(\vec{k}_{1})+
\delta S_{2}(\vec{k}_{1}) \delta S_{3}(\vec{k}_{2}))\nonumber
\\ \times [\delta_{\vec{k}_{1}+\vec{k}_{2}
+\vec{q}_{1}+\vec{q}_{2},0}-\delta_{\vec{k}_{1}+\vec{k}_{2}-
\vec{q}_{1}-\vec{q}_{2},0}+\delta_{\vec{k}_{1}+\vec{k}_{2}+
\vec{q}_{1}-\vec{q}_{2},0}-\delta_{\vec{k}_{1}+\vec{k}_{2}+
\vec{q}_{2}-\vec{q}_{1},0}]] \} \nonumber
\\ + \frac{2 u a_{1}^{2}}{N} \sum_{\vec{k}_{1},\vec{k}_{2}} 
\{  \delta S_{1}(\vec{k}_{1})\delta S_{1}(\vec{k}_{2}) 
[\delta_{\vec{k}_{1}+\vec{k}_{2},0}+\frac{1}{2}
(\delta_{\vec{k}_{1}+\vec{k}_{2}+2 \vec{q}_{1},0}+
\delta_{\vec{k}_{1}+\vec{k}_{2}-2\vec{q}_{1},0})] \nonumber
\\+ \delta S_{2}(\vec{k}_{1}) \delta S_{2} (\vec{k}_{2}) 
[ \delta_{\vec{k}_{1}+\vec{k}_{2},0} - \frac{1}{2}
(\delta_{\vec{k}_{1}+\vec{k}_{2}+2 \vec{q}_{1},0}
+\delta_{\vec{k}_{1}+\vec{k}_{2}-2 \vec{q}_{1},0})] \}
\nonumber
\\ +\frac{2 u a_{2}^{2}}{N} \sum_{\vec{k}_{1},\vec{k}_{2}} 
\{ \delta S_{3}(\vec{k}_{1}) \delta S_{3}(\vec{k}_{2}) 
[\delta_{\vec{k}_{1}+\vec{k}_{2},0} +\frac{1}{2} 
(\delta_{\vec{k}_{1}+\vec{k}_{2}+2 \vec{q}_{2},0}+
\delta_{\vec{k}_{1}+\vec{k}_{2}-2\vec{q}_{2},0})] \nonumber
\\+ \delta S_{4}(\vec{k}_{1}) \delta S_{4}(\vec{k}_{2}) 
[\delta_{\vec{k}_{1}+\vec{k}_{2},0} -\frac{1}{2} 
(\delta_{\vec{k}_{1}+\vec{k}_{2}+2\vec{q}_{2},0}+
\delta_{\vec{k}_{1}+\vec{k}_{2}-2\vec{q}_{2},0})]\} \nonumber
\\ -\frac{i u a_{1}^{2}}{N} \sum_{\vec{k}_{1},\vec{k}_{2}} 
[\delta S_{1}(\vec{k}_{2}) \delta S_{2}(\vec{k}_{1}) +\delta S_{1}(\vec{k}_{1})
 \delta S_{2}(\vec{k}_{2})](\delta_{\vec{k}_{1}+\vec{k}_{2}+2\vec{q}_{1},0}
-\delta_{\vec{k}_{1}+\vec{k}_{2}-2\vec{q}_{1},0}) \nonumber
\\ -\frac{i u a_{2}^{2}}{N} \sum_{\vec{k}_{1},\vec{k}_{2}} 
[\delta S_{3}(\vec{k}_{2})\delta S_{4}(\vec{k}_{1})
+ \delta S_{3}(\vec{k}_{1}) \delta S_{4}(\vec{k}_{2})]
(\delta_{\vec{k}_{1}+\vec{k}_{2}+2\vec{q}_{2},0}
-\delta_{\vec{k}_{1}+\vec{k}_{2}-2 \vec{q}_{2},0}).
$

We may obtain an equation for 
the eigenvalues by truncating the expansion 
for $\det [{\cal{H}}-E]$ at $O(u^{2})$.
Just as the dispersion relation for low lying states
could have been derived by setting $\det [{\cal{H}} -E]=0$
to $O(u^{2})$ and solving the resultant quadratic
equation for the low lying energy states:
$\vec{k} = \vec{q}+ \vec{\delta}$ with small $\vec{\delta}$. 
Here we may do the same. 
The characteristic equation
is easily read off: 
\begin{eqnarray}
0=\det[{\cal{H}}-E]= 
\prod_{i=1}^{N} [(v(\vec{k}_{i}) - E)^{2}] - \nonumber
\\ 
4u^{2} a_{1}^{2} \sum_{j_{1}} [(v(\vec{k}_{j_{1}})+8u)
(v(\vec{k}_{j_{1}}+2 \vec{q}_{1}) +8u)]\nonumber
\\ \times \prod_{i \neq j_{1}} [(v(\vec{k}_{i})-E]^{2}] \nonumber
\\ - 4u^{2}a_{2}^{2} \sum_{j_{2}} [(v(\vec{k}_{j_{2}})+8u)
(v(\vec{k}_{j_{2}}+2 \vec{q}_{2}) +8u)]  
\nonumber
\\ \times  \prod_{i \neq j_{2}} [(v(\vec{k}_{i})-E]^{2}]
\end{eqnarray}
where we have shifted $E$ by a constant.
Notice that decoupling trivially occurs
- terms of the form $[v(\vec{k})-E)][v(\vec{k}^{\prime})-E]$
where the modes 
$\vec{k}-\vec{k}^{\prime} = \vec{q}_{1} 
\pm \vec{q}_{2}$, cancel.
The four coupled polarizations break into
two pairs and that $H_{1}$
plays the role of a pairing interaction.
Schematically, for a high dimensional minimizing manifold $M$, 
with low lying states i.e.
for the terms containing 
$\vec{k}_{1} = \vec{q}_{1} + \vec{\delta}_{1}$
and $\vec{k}_{2} = \vec{q}_{2} + \vec{\delta}_{2}$:
\begin{eqnarray}
 E= E_{\min} + a_{1}^{2} [A_{||} \delta_{||;1}^{2} +
 A_{\perp} \delta_{\perp;1}^{4}] \nonumber
\\ 
+ a_{2}^{2} [A_{||} \delta_{||;2}^{2} + A_{\perp} \delta_{\perp;2}^{4}]
\end{eqnarray}
with $\delta_{||,\perp;m}$ parallel and 
perpendicular to the minimizing manifold $M$ 
at $\vec{q}_{m}$, 
trivially satisfies $\det[{\cal{H}}-E]=0$.


This dispersion relation agrees, once again, with the result 
derived from the spin wave stiffness analysis
\begin{eqnarray}
\Delta E = \frac{1}{2N}\sum_{\vec{k}} (v(\vec{k})- v(\vec{q}))
|\vec{S}(\vec{k})|^{2}
\end{eqnarray}
and by expansion of $\Delta H$ for different sorts of twists,
the dispersion relations of the two spiral
simply lumped together. When $d=3$, as in the 
$O(2)$ case ($p=1$)
this dispersion gives
rise (in the Gaussian approximation)
to diverging logarithmic
fluctuations: $O(| \ln ~\epsilon|)$.
Applying equipartition, the Gaussian spin
fluctuations in the $[2i-1,2i]$ plane:
\begin{eqnarray}
\Delta \vec{S}^{2}_{[2i-1,2i]}(\vec{x}=0) > \Delta S^{2}_{low~~[2i-1,2i]}
(\vec{x}=0)
\nonumber \\ = k_{B} T
\int \frac{d^{3}k}{(2 \pi)^{3}}~~ \frac{1}{a_{i}^{2} [A_{||} \delta_{||;i}^{2}
+ A_{\perp} \delta_{\perp;i}^{4}]}.
\end{eqnarray} 

As the spin fluctuations in the 
[12] and [34] plane decouple,
we may easily compute the 
correlation functions to
find that they are a sum
of two decoupled pieces
of the form (Eq.(\ref{ALRO}))
found earlier for XY systems.
For general rotationally 
symmetric incommensurate spin
systems having a shell of 
minimizing modes of radius $q>0$ in
momentum space, fluctuations about a certain 
polyspiral state explicitly lead to 
the soft-spin
\begin{eqnarray}
G(\vec{x}) 
\simeq \sum_{i=1,2} 
\frac{4 a_{i}^{2} d^{2}\cos(qx_{i;||})}{x_{\perp}^{2}}  \exp[- 2 \eta \gamma -
\eta E_{1}(\frac{x_{i;\perp}^{2} Q^{1/4}}{2 x_{i;||}})],
\nonumber
\end{eqnarray}
where $x_{i,||}$ denotes the spatial coordinate
in a direction parallel to $\vec{q}_{i}$ etc. 
As before, $\eta= \frac{k_{B}T}{16 \pi}Q^{1/4}$, $d= \frac{2 \pi}{\Lambda}$ 
where $\Lambda$ is the  
ultra violet momentum cutoff, $\gamma$ is Euler's const., $q=Q^{1/4}$, and 
$E_{1}(z)$ is the exponential integral.

For all odd $n$ and for all even $n$ with $p<n/2$ there
will be divergent fluctuations similar to those encountered
for the $O(3)$ model, no algebraic long range order
is found. 

In conclusion,
if frustrating interactions
cause the ground states to be modulated
then the associated  ground state
degeneracy (for $n>2$) is much
larger by comparison to the 
usual ferromagnetic ground states.
For even $n$ we 
have found that, generically, 
the a three dimensional system
will not have long range order
when $M$ is two dimensional.
When $n$ is odd the system will
never show long range order
if $M$ is (d-1) or (d-2) dimensional.
If, in the continuum limit, the uniform
(ferromagnetic) state is higher
in energy than any other state
then, by rotational symmetry,
the manifold of minimizing modes in Fourier 
space is $(d-1)$ dimensional.
In reality, small symmetry breaking terms (e.g. $\lambda \neq 0$
in $v_{Q}(\vec{k})$) will 
always be present- these will favor ordering at a 
discrete set of $\{ \pm\vec{q}_{m}\}_{m=1}^{|M|}$.
If $n>2|M|$ then, irrespective of the
even/odd parity of $n$, then will be a
 \begin{eqnarray}
\langle \Delta \vec{S}^{2}(\vec{x}=0) \rangle \ge
(n- 2 |M|)\int \frac{d^{d}k}{(2 \pi)^{d}}
\frac{k_{B}T}{v(\vec{k})-v(\vec{q})}.
\label{leftover}
\end{eqnarray}
This contribution is monotonically increasing in $n$; Within our
scheme, $T_{c}$ is finite and may be estimated by
the temperature at which the fluctuations are of
order unity. By tweaking the symmetry breaking
terms to smaller and smaller values, the 
fluctuation integral becomes larger and larger.
For instance, if take $\lambda \ll 1$ 
in $v_{Q}(\vec{k})$ then the integral
is very large and $T_{c}$ extremely low
(in can be made arbitrarily low).
Thus as the system will be cooled from 
high temperatures, it might
first undergo a Kosterlitz-Thouless 
like transition at $T_{KT}$ to an 
algebraically ordered state and
develop true long range order 
at critical temperatures $T_{c}<T_{KT}$.

\section{The Large n limit- Fluctuation spectrum, ground
states, ground state entropy, and $T_{c}$}
\label{spherical}

So far we have seen that the classical even $n$
systems are more ``gapped'' than their
odd counterparts. This is reminiscent of the appearance/non-appearance
of a gap for quantum spin $s$ chains with even/odd values of $(2s+1)$ 
\cite{Haldane}, \cite{Fradkin} and the physics for an 
even/odd number of legs in 
an $s=1/2$ quantum spin ladder \cite{sudip} where
such even/odd discrepancies are triggered by Berry phase terms.
There is no problem in the classical large $n$ (or spherical
model) limit (just as there is none in the quite
different quantum spin systems). 
In this limit, wherein a single normalization
constraint is imposed
\begin{eqnarray}
\sum_{\vec{x}} S^{2}(\vec{x}) = N,
\label{N-glob}
\end{eqnarray}
the effective number of spin components
$n$ is of the order of the number of 
sites in the system $N$. The span
of the system $N$ (the number
of Fourier modes allowed within the 
Brillouin zone) is always larger
than the number of minimizing modes
$\{\vec{q}_{i}\}$. 
In such a case we will be 
left with a divergence as
in equation (\ref{leftover})
due to the many unpaired
spin components.
Within the spherical 
model, which is easily
solvable the fluctuation
integral exactly marks the 
value of the inverse critical
temperature
\begin{eqnarray}
\frac{1}{k_{B}T_{c}} = \int_{B.Z.} \frac{d^{d}k}{(2 \pi)^{d}} 
~ ~ \frac{1}{v(\vec{k})-v(\vec{q})}.
\end{eqnarray}
Thus, $T_{c} =0$ if the latter integral
diverges and our circle of ideas nicely 
closes on itself. In terms of ground state degeneracies, 
in the large $n$
limit, any configuration
\begin{eqnarray}
S^{g}(\vec{x}) = \sum_{\vec{q} \in M} S(\vec{q}) e^{i \vec{q} \cdot \vec{x}}
\end{eqnarray}
satisfying the ``reality'' condition ($S(\vec{x}) = S^{*}(\vec{x})$) in
momentum space  
$S^{*}(\vec{q}) = S(-\vec{q})$,
and global normalization (Eq.(\ref{N-glob}))
\begin{eqnarray}
\sum_{\vec{q} \in M} |S(\vec{q})|^{2} = N^{2}
\end{eqnarray}
is a ground state. The proof
is the same as before: any configuration
$S^{g}(\vec{x})$ having as its non-vanishing Fourier
components only those momenta
$\vec{q}$ that minimize the interaction
kernel $v(\vec{k})$, is a ground
state. Thus, with the 
ground state degeneracy denoted 
by $g$, 
\begin{eqnarray}
\ln(g) \sim |M|.
\end{eqnarray}
The ground state entropy is
given by the number of minimizing
modes in $\vec{k}$ space- 
the size of the manifold 
$M$ spanned by these
modes. As the spherical
model is the least restricted 
of all $O(n)$ variants,
this serves as an
upper bound on 
finite $n$, $O(n)$
ground state degeneracies.
The astute reader
will note that the Ising
ground state constructed 
in section (\ref{Ising-gs})
for the kernel $v_{z} = z \Delta^{2} - \Delta$
with $z=1/8$
on the square lattice indeed do
saturate the spherical bound.
In the situation that the minimizing
modes lie on $(d-1)$ dimensional
manifold $M$, the zero temperature
entropy is sub-extensive yet very large, 
scaling with the surface area of the 
system- $\ln(g) \sim L^{d-1}$.
We note that 
arbitrary periodic patterns
(including amongst many
others Wigner crystals)
will be found in 
the spherical re-incarnation
of many spin models \cite{wigner}.

\section{Permutational
Symmetry and the Integrability
of the Spherical Model}
\label{permutational}

In many integrable models, 
the system possesses an
extremely large symmetry in
the thermodynamic limit.
Perhaps closest in  
spirit to the 
spherical model
which possesses
an $N!$ fold permutational
symmetry is a permutational
like character and symmetry
within integrable Bethe
ansatz models where not only
the total momentum
is conserved but
also the momentum
of each individual particle 
does not change apart from
permutations upon
scattering.
Although trivial,
the permutational symmetry
of the large $n$ model
has often been overlooked.
In fact, several
authors have attempted
(unsuccessfully)
to find systems
having a different 
geometry of minimizing
manifolds in $\vec{k}-$
space (spherical
surfaces versus sheets
etc.) yet with the
same degeneracy that
have different 
thermodynamics.
This quest was 
unsuccessful
for fundamental
reasons. The classical spin spherical model (or  $O(n \rightarrow \infty)$) 
partition function
\begin{equation}
  Z = const \left(\prod_{\vec{k}}
    \left[\frac{1}{\sqrt{\beta[\hat{v}(\vec{k})+\mu]}}\right]\right),
\end{equation}
where the chemical
potential
$\mu$ satisfies
\begin{eqnarray}
\beta = \int \frac{d^{d}k}{(2 \pi)^{d}} \frac{1}{\hat{v}(\vec{k}) + \mu},
\end{eqnarray}
is invariant under permutations of $\{\hat{v} ( \vec{k})\} \rightarrow
\{\hat{v}(P\vec{k})\}$. In the above, the permutations
\begin{eqnarray}
\{\vec{k}_{i} \}_{i=1}^{N} \rightarrow \{ P \vec{k} \}
\end{eqnarray}
correspond to all possible shufflings of the 
$N$ wave-vectors $\vec{k}_{i}$. Although quite simple,
this is not universally realized. Several authors
have attempted to compute the critical exponents
in the spherical limit (via an RG calculation)
for systems having different minimizing manifolds
yet all sharing the same relevant density of states.
This quest was not very economical. As unrealized by 
these authors, by permutational symmetry these models are identical.

This simple invariance allows all d-dimensional 
translationally invariant systems to 
be mapped onto a 1-dimensional one. Let us design an effective
one dimensional kernel $V_{eff}(k)$ by
\begin{equation}
  \int \delta[\hat{v}(\vec{k})-v] d^{d}k =
  |\frac{dV_{eff}}{dk}|_{V_{eff}(k)=\hat{v}}^{-1}.
\end{equation}
The last relation secures that
the density of states and consequently
the partition function is preserved.  For 
the two-dimensional nearest-neighbor ferromagnet:
\begin{eqnarray}
  |\frac{dV_{eff}}{dk}|^{-1} =\rho(V_{eff}) \nonumber
\\ = c_{1}
  \int_{0}^{1}\frac{dx}{\sqrt{1-x^{2}}\sqrt{1-(V_{eff}+x-2)^{2}}},
\end{eqnarray}
and consequently 
\begin{eqnarray}
  k(V_{eff})= c_{1}\int_{0}^{V_{eff}}
  F(\sin^{-1}\sqrt{\frac{2}{(3-u)u}},\nonumber
\\ \frac{\sqrt{4u- u^{2}}}{2}) du,
\label{elliptic}
\end{eqnarray}
where $F(t,s)$ is an incomplete elliptic integral of the first kind.
Eq.(\ref{elliptic}) may be inverted
and Fourier transformed to find the effective one dimensional 
real space kernel $\hat{V}_{eff}(x)$.
We have just mapped the two dimensional nearest 
neighbor ferromagnet onto a one dimensional 
system.
In a similar fashion, within the spherical 
(or equivalently the $O(n \rightarrow \infty)$) limit
 all high dimensional problems may be mapped
onto a translationally invariant one dimensional 
problem. It follows that 
the, large $n$, critical exponents of the
$d$ dimensional nearest neighbor ferromagnet
are the same as those of translationally invariant 
one dimensional system with longer range
interactions. We have just shown 
that a two dimensional $O(n \gg 1)$ 
system may has the same thermodynamics
as a one dimensional system.
By permutational symmetry,
such a mapping may be performed
for all systems irrespective
of the dimensionality of the 
lattice or of the nature of the 
interaction (so long
as it translationally
invariant). This demonstrates once again
that the notion of universality 
(with dependence only on the order parameter symmetry, 
dimensionality etc.) may apply only to the canonical 
interactions.

Permutational symmetry is broken 
to ${\cal{O}}(\beta^{4})$ for finite $n$.
Upon performing
a Hubbard-Stratoniwich transformation,
for a constraining term (e.g. $\sum_{\vec{x}} \ln[\cosh[\eta(\vec{x})]]$ for
$O(1)$ spins) symmetric in $\{\eta(\vec{k})\}$ to a given order, one
may re-arrange the non-constraining term $\sum_{\vec{k}}
\hat{v}^{-1}(\vec{k}) |\eta(\vec{k})|^{2} = 
\sum_{\vec{k}} \hat{v}^{-1}(P \vec{k})
|\eta(P \vec{k})|^{2}$) and relabel the dummy integration variables
$H[\{\eta(\vec{k})\}] \rightarrow H[\{\eta(P^{-1}\vec{k})\}]$ to effect
the constraining term augmented to a shuffled spectra $\hat{v}
(P \vec{k})$.

\section{$O(n \ge 2)$ Weiss Mean Field Theory
0f Any Translational Invariant Theory.}
\label{high_mft}

In this section the critical temperature is
computed exactly for all translationally invariant
incommensurate $O(n \ge 2)$ systems. 
Unlike the situation for Ising spins,
we proved that simple spiral
states are the only ground
states for incommensurate systems. Here, 
for $O(n \ge 2)$ when $T<T_{c}$:
\begin{equation}
\langle \vec{S}(\vec{x}) \rangle ~ = ~s ~ \vec{S}^{ground-state}(\vec{x}).
\end{equation}
For the particular case
\begin{eqnarray}
S_{1}^{ground-state}(\vec{x}) = \cos(\vec{q} \cdot \vec{x}) \nonumber
\\ S_{2}^{ground-state}(\vec{x}) = \sin(\vec{q} \cdot \vec{x}) \nonumber
\\ S_{i \ge 3}^{ground-state}(\vec{x}) =0 
\end{eqnarray}
Now only the $ \pm \vec{q}$ modes have finite weight. 
Repeating the previous steps of section(\ref{ISING-MFT}), 
\begin{eqnarray}
\sum_{\vec{y}} V(\vec{x}=0,\vec{y}) S_{2}^{ground-state}(\vec{y}) = 0 
\end{eqnarray}
and $| \langle \vec{S}(\vec{x}=0) \rangle | = | \langle
S_{1}(\vec{x}=0) \rangle | \hat{e}_{1}$.
We now define 
\begin{eqnarray}
M[z] \equiv - \frac{d}{dz}  \ln [(2/z) ^{(n/2-1)} I_{n/2-1}(z)],
\end{eqnarray} 
with $[I_{n/2-1}(z)]$ a Bessel function.
With this definition in hand, the mean-field equation reads 
\begin{eqnarray}
|\langle S_{1}(\vec{x}=0) \rangle | = s = M[~|\sum_{\vec{y}}
V(\vec{x},\vec{y}) \langle S_{1}(\vec{y}) \rangle| ~].
\nonumber
\end{eqnarray}
The onset of the non-zero solutions is at
\begin{equation}
|\beta_{c} v(\vec{q})| = n.
\end{equation}
If $V(\vec{x}=0)=0$ (no on-site interaction), then 
$\int d^{d}k~ v(\vec{k}) =0$, 
implying that $v(\vec{q}) < 0 $
and $T_{c}>0$. Note that within the
mean field approximation,
$T_{c}$ is a continuous function
of all the parameters in the 
Hamiltonian.
Here the ground state is symmetric with respect to all sites.
The above is the exact value of $T_{c}$ 
within Weiss mean field theory for the helical ground-states.  
For poly-spirals we will get $p$ identical equations:
both sides of the self consistency equations
are multiplied by $a_{l}^{2}$ where
$a_{l}$ is the amplitude of the $l-th$ spiral in the
[$(2l-1),2l$] plane. As $v(\vec{q}_{m}) = v(\vec{q})$,
we will arrive at the same value of $T_{c}$ as for the
case of simple spirals.

\section{Extensions to arbitrary two spin interactions}
\label{brilliant}

Any real kernel $V(\vec{x},\vec{y})$ may be symmetrized 
$[V(\vec{x},\vec{y}+ V(\vec{y},\vec{x})]/2 \rightarrow
V(\vec{x},\vec{y})$ to a hermitian form. 
Consequently, by a unitary transformation,
it will become diagonal. The Fourier 
modes are the eigen-modes of $V$ when
it is translationally invariant.
We may similarly envisage extensions
to other, arbitrary, $V(\vec{x},\vec{y})$  
diagonalized in
another complete orthogonal basis $\{ |\vec{u} \rangle \}$:
\begin{eqnarray}
\langle \vec{u}_{i} | V | \vec{u}_{j} \rangle =  \delta_{ij}  
\langle \vec{u}_{i} | V | \vec{u}_{i} \rangle
\end{eqnarray}

Many of the statements that we have made hitherto
have a similar flavor in this more
general case.

For instance, the large $n$ fluctuation
integrals are of the same form
\begin{eqnarray}
\int d^{d}u~ ~ \frac{1}{v(\vec{u}) - v_{\min}}
\end{eqnarray}
with the wave-vector $\vec{k}$ 
traded in for $\vec{u}$. Here, $v_{\min} \equiv \min_{\vec{u}} v(\vec{u})$.  
Once again, one may examine the topology 
of the minimizing manifold in $\vec{u}$
space. If the surface if $(d-1)$ dimensional
and $v(\vec{u})$ is analytic in its
environs then, for large $n$, 
$T_{c} =0$. More generally, in the 
large $n$ limit, the 
partition function 

\begin{equation}
  Z = const \left(\prod_{\vec{u}}
    \left[\frac{1}{\sqrt{\beta[v(\vec{u})+\mu]}}\right]\right),
\end{equation}
where the chemical
potential
$\mu$ satisfies
\begin{eqnarray}
\beta N = \sum_{\vec{u}} \frac{1}{v(\vec{u}) + \mu}.
\end{eqnarray}

The topology of the ground state sector
of $O(n)$ models will once again be 
governed by a direct product of
the topology of the minimizing manifold
in $\vec{u}$ space with the 
spherical manifold of the 
$O(n)$ group. In the general
case it will be dramatically
rich.

We may similarly extend the Peierls
bounds of section(\ref{peierls}) 
to some infinite range interactions
also in this case by contrasting
the energy penalties in the now
diagonalizing $\vec{u}$
basis with those that occur
for short range systems which are 
diagonal in 
Fourier space. For instance, we 
may examine spin configurations computed
with the non-translationally invariant (``disordered'')
kernel $V$ and compare the energies relative
to those in a simple reference short range interaction 
which is diagonal in $k-space$ ($v_{short}(\vec{k})$) and 
has a minimum $v_{short ~ \min} \equiv \min_{\vec{k}} \{v_{short}(\vec{k})\}$.
The energy penalties for excitations in a system with the 
kernel $V$,  
\begin{eqnarray}
\Delta E_{disordered} = \frac{1}{2N} 
\sum_{\vec{u}} |\vec{S}(\vec{u})|^{2} 
[\langle \vec{u} | V| \vec{u} \rangle - v_{min}(\vec{u})]
\nonumber
\\ 
\ge \frac{1}{2N} \sum_{\vec{k}} 
|\vec{S}(\vec{k})|^{2} [v_{short}(\vec{k}) - v_{short~\min}]
\end{eqnarray}
if the kernels $V$ and $V_{short}$
share the same realizable lowest energy 
Ising eigenstate.

\section{O(n) spin dynamics and simulations}
\label{dynamics}

We now investigate the dynamics in all
(commensurate or incommensurate) continuous
spin systems. It is seen that the 
equations of motion are relatively
trivial in $\vec{k}$ space. Potentially,
this allows for the construction
of new efficient algorithms. If the Hamiltonian
\begin{eqnarray}
H = \frac{1}{2} \sum_{\vec{x},\vec{y}} V(\vec{x},\vec{y}) 
\vec{S}(\vec{x}) \cdot \vec{S}(\vec{y}),
\end{eqnarray}
then the force, 
$\vec{F}(\vec{z}) = -\frac{\partial H}{\partial \vec{S}(\vec{z})}$,
on given spin at size $\vec{z}$ is 
\begin{eqnarray}
\vec{F}(\vec{z}) = 
- \frac{1}{2} \sum_{\vec{y}} [V(\vec{z},\vec{y}) 
+ V(\vec{y},\vec{z})]  \vec{S}(\vec{y}).
\end{eqnarray}
We may symmetrize $[V(\vec{x},\vec{y}+ V(\vec{y},\vec{x})]/2
\rightarrow V(\vec{x},\vec{y})$, as we have done repeatedly, 
without changing $H$. 
For a more general two spin kernel 
$V(\vec{x},\vec{y})$ which is not translationally
invariant the Fourier space index $\vec{k}$
should be replaced by the more general $\vec{u}$.  
It is readily verified that 
\begin{eqnarray}
\frac{d^{2}\vec{S}(\vec{k})}{dt^{2}} = 
- \frac{1}{N}v(\vec{k}) \vec{S}(\vec{k}).
\end{eqnarray}
Let an arbitrary constant $A$ satisfy
$A > - \min_{\vec{k}} \{ v(\vec{k}) \}$.
The equations of motion in a shifted system
with the interaction kernel $(v(\vec{k}) + A)$ are 
\begin{eqnarray}
\frac{d^{2}\vec{S}(\vec{k})}{dt^{2}} = -[A+v(\vec{k})] \vec{S}(\vec{k})
\nonumber
\\ \equiv -\omega_{k}^{2} \vec{S}(\vec{k}),
\label{eqn_motion}
\end{eqnarray}
which may be trivially integrated. Adding the constant $A$ merely shifts 
$H \rightarrow H+ A/2$ with no change
in the underlying physics. The solution to Eq.(\ref{eqn_motion}) is
\begin{eqnarray}
\vec{S}_{un}(\vec{k},t) = \vec{S}(\vec{k},0) \cos \omega_{k} t 
+ \frac{d \vec{S}(\vec{k},t)}{dt}|_{t=0}~ \times  \omega_{k}^{-1} 
\sin \omega_{k} t, \nonumber
\\
\vec{S}_{un}(\vec{k}, t+ \delta t) = \vec{S}(\vec{k},t) \cos
\omega_{k} \delta t \nonumber
\\ + \frac{\delta \vec{S}(\vec{k},t)}{\delta t} ~ \omega_{k}^{-1} \sin
\omega_{k} \delta t.
\nonumber
\end{eqnarray}
This suggests the following simple recursive algorithm:
(i)   At time t, start off with initial values $\{\vec{S}(\vec{x}, t) \}$;
(ii)  Fourier transform to find $\{ \vec{S}(\vec{k},t) \}$;
(iii) Integrate to find the un-normalized 
$\{ \vec{S}_{un}(\vec{k},t+\delta t) \} $;
(iv)  Fourier transform back to find the un-normalized real-space spins
$\{ \vec{S}_{un}(\vec{x},t+\delta t) \} $; 
(v) Normalize the spins: 
\begin{eqnarray}
\vec{S}(\vec{x},t+\delta t) = 
\frac{\vec{S}_{un}(\vec{x},t+\delta t )}{|\vec{S}_{un}(\vec{x},t+\delta t)|};
\end{eqnarray}
(vi)  Compute $ \{ \delta \vec{S}(\vec{x}, t+\delta t) \} = 
\{ [\vec{S}(\vec{x},t+\delta t)-\vec{S}(\vec{x},t)] \} $;
(vii) Fourier transform to find $\{ \delta \vec{S}(\vec{k},t+\delta t)\}$;
and finally (viii) Go back to (ii).
Thus far, we neglected thermal effects. 
To take these into account, we could integrate these equations
with a thermal noise term augmented to the 
restoring force 
\begin{eqnarray}
\frac{d^{2}\vec{S}(\vec{k})}{dt^{2}} = - \frac{1}{N} v(\vec{k}) 
\vec{S}(\vec{k}) + \vec{F}_{\vec{k}}^{~noise}(T,t).
\end{eqnarray}
Expressed in this format, the execution of this algorithm for continuous
$O(n \ge 2)$ spins seems easier than that for a discrete Ising system.
Here the equations of motion may be integrated to produce arbitrarily
small updates at all sites. A prioi, this algorithm might be better 
than a real space 
more brute force approach  whereby, effectively, the torque 
equations in angular variables
(the spins $\vec{S}(\vec{x})$ are automatically normalized)
are integrated \cite{eqs_of_motion}.

\section{Conclusions}

In this article, we discussed 
$O(n)$ spin models on a cubic lattice 
in mostly translationally invariant systems.
These include frustrated
systems that have been the subject
of much attention in recent years.
Here we outline a few of 
our results.

{\bf 1}) Certain frustrated Ising systems were shown to
display a huge ground state degeneracy. The 
ground state entropy in these systems
scaled as the surface area, illustrating
a ``holographic'' like effect.
Computing the energies of 
various contending states
is trivial and easily
allows us, on a non-rigorous
footing, to see if lock-in
effect may arise and what
scaling might be anticipated.

{\bf 2}) The Peierls energy bound
may be rigorously extended 
to many Ising systems 
in which the interaction kernel displays 
minima at a finite number of 
commensurate wave-vectors. These 
systems include some with arbitrarily
long range interactions. This bound can 
also be extended in certain instances to 
systems displaying no translational
invariance.

{\bf 3})  The ground states of 
all translationally invariant XY 
models were rigorously shown
to be spirals (and only spirals) unless certain
special commensurability 
conditions were met. No
other ground state are 
generically possible.
The trivial uniform 
and Neel states
correspond to commensurate
wave-number ordering 
(or spirals of infinite 
or two unit length pitch
respectively). 

{\bf 4}) The Weiss mean-field equations
of the general $O(n)$ systems were looked
at. The mean field transition temperature 
$T^{MF}$ for all $O(n \ge 2)$ models
was exactly computed.

{\bf 5}) For $O(n \ge 3)$ spins, 
poly-spiral states are the dominant ones.
These correspond to a hybrid of 
$p \le Int [n/2]$ (with $Int[ ] $ denoting the
integer part) spirals in different
orthogonal planes.  Once again, unless
commensurability conditions amongst
the minimizing modes are 
met, we can easily prove
that these are the only viable
ground states. It is also easy to demonstrate
that the poly-spiral states
with the largest viable 
$p$ (i.e. $p= Int [n/2]$)
are statistically preferred.
Moreover, they are more stable
against thermal fluctuations.

{\bf 6}) We performed a non-rigorous spin stiffness 
analysis to test the stiffness of XY spins 
under different sorts of external twists. 
We found that certain frustrated
systems display a smectic like 
low energy dispersion.

{\bf 7}) A thermal fluctuation
analysis of soft XY spins for arbitrary interactions
may be easily done. We found that, in the general case, a Dirac like equation 
arises for the two component ``spinors''. In general,
for incommensurate momenta, the different momenta are coupled.
By truncating the equations for 
small $u$ (the soft spin limit),
we regained the exact same energy 
dispersion attained by the simple
minded spin-stiffness analysis. 
We found that in rotationally
symmetric incommensurate
spin systems,
the resulting spectra matches that of
smectic liquid crystals. Liquid crystalline
properties are natural for many
of the frustrated systems appearing
when competing interactions
are present.

{\bf 8}) We generalized the Mermin-Wagner-Coleman
theorem to all translationally invariant
$d \le 2$ systems with a twice differentiable
interaction kernel. Here we did not only
focus on ferromagnetic or antiferromagnetic
states. All possible orders were excluded.
The connection of the denominator to
Gallilean boosts was noted. At zero temperature,
the denominator of the integrand in 
the general inequality matches that
attained by the Dirac like analysis.

{\bf 9}) The resulting
generalized Mermin-Wagner-Coleman 
bounds may be examined in high dimensions ($d>2$).
We showed that if a certain integral
diverges, the generalized Mermin-Wagner-Coleman
bounds disallow us to think of the dispersion 
relation for spin waves at arbitrarily low 
temperatures $T=0^{+}$ as that computed by 
assuming a perfectly ordered 
ground state. We also formally
linked the integrand to 
a decoherence (or bandwidth) time scale
\cite{moessner} in the
quantum case. We further noted
that exactly such a divergence
of ``decoherence time'' is linked
to a non-rigorous replica derivation of glassy
behavior in many systems. 
A large degeneracy and 
near degeneracy may 
be the underlying cause of
two effects- a greater fragility
of order and sluggish 
glassy dynamics.
	
{\bf 10}) The 
effect of thermal fluctuation
in the most general
translationally invariant 
$O(n=3)$ (Heisenberg) systems was 
considered. We showed that,
for incommensurate wave-numbers, 
the effects of fluctuations 
here are expected to be larger
than in XY spins. The only
approximation made here
was that the spin fluctuations
$\delta S$ about the only viable 
ground state (proved
in an earlier section)
were small. This in turn
led to a paradox at finite
temperature: assuming only small
spin fluctuations led
to a divergent thermal
fluctuation disallowed
by the normalization of
the spin. The parameter
$u$ enforcing the normalization
of the spin via the additional
term $H_{1}= u \sum_{\vec{x}} (\vec{S}^{2}(\vec{x})-1)^{2}$,
may be arbitrarily large in
this treatment (we are not
confined to only
soft spins $u \ll 1$). The only
assumption made in this
non-rigorous analysis
that $|\delta \vec{S}| \ll 1$.
For most canonical
systems with kernels analytic about point minima 
in $\vec{k}$ space (e.g. the nearest neighbor
Heisenberg model), this integral converges
and estimated $T_{c}$ is sizable. However, for the 
Coulomb frustrated
model and several others,
this was seen to suggest
that $T_{c}$ might
be zero as the corresponding
fluctuation integral diverges
in that case.  

{\bf 11}) We extended our thermal
fluctuation analysis
to all translationally invariant $O(n \ge 4)$ systems.
For even $n$, the determination
of the spectrum led to the analysis
of the Dirac like equation obtained
for the XY model. For odd $n$,
one component was left unpaired
and the fluctuations were seen
to be much larger. 
The integral appearing 
in the generalized Mermin-Wagner 
inequality 
is a strict lower bound
on the fluctuations.
Thus, spin systems with a non-trivial minimizing manifold 
in $\vec{k}$ space, much like spin ladders, display an 
interesting {\em even-odd effect}. When we analyze the
fluctuations of spins having an
even number of components (n),
we find precisely the Mermin-Wagner integral
as the relevant thermal fluctuation integral.
For a system of spins having an 
odd number of components, we find 
a more divergent (for on-shell
minima in Fourier space) fluctuation integral.
Under the 
influence of the quartic $H_{1}$ term,
spin (or field) components
bunch up in pairs. For an odd
number of spin components $n$, a
lone unpaired spin 
component can give rise to
more divergent fluctuations.
Computing the correlators within truncated 
soft spin models, we
found that, in rotationally symmetric
incommensurate systems, {\em weak algebraic 
long range order may persist in even $n$ systems,
while being absent in odd $n$ systems of the
same Hamiltonian}.
Much as in quantum spin ladders
and chains,
the energy spectrum 
is more ``gapped'' and inhibited
for even spins.
In the large $n$ limit,
both tend to the ``odd'' $n$
spherical limit fluctuation
integral which is none
other than the spherical limit
expression. The bottom line is that the bound
that we derive by a generalized 
Mermin-Wagner inequality
is a strict lower bound
on the fluctuations. For
odd component spins $n$, the
thermal fluctuations are much 
larger. This goes against 
some of the common held
intuition  (correct
for conventional commensurate ordering at
$\vec{q} = 0, (\pi,\pi, ..., \pi)$)
that as $n$
is monotonically
increased, the 
system orders at 
lower and lower
temperatures as 
entropy effects
become ever large.
By contrast,
we find that
for incommensurate
orders, the size of
the thermal fluctuations
is not monotonic
in $n$ but rather
exhibits novel even-odd
alterations.

{\bf 12}) The large $n$ limit,
which is exactly solvable,
gives a fluctuation integral
($1/T_{c}$) similar to that
for the odd $n$ case. 
We further rigorously
illustrate how the ground
state entropy scales as the
size of the manifold of minimizing
modes in $\vec{k}$ space. 
In particular, for many
frustrated models, this
leads to an entropy which
increases as the surface area
of the system, leading to 
a {\em ``holographic''} like effect.

{\bf 13}) We illustrated how 
we may think
about the topology of the
minimizing manifold and
use other concepts that
we introduced hitherto 
also when the interaction
is not translationally invariant,
as in disordered systems.
In these cases we considered the
new coordinate $\vec{u}$ parameterizing
the eigenstates of the interaction
kernel.  The large $n$ 
analysis proceeds as before.
We illustrated
how Peierls' bounds 
may be constructed 
in certain systems
with long range
interactions.

{\bf 14}) The dynamics
of $O(n)$ spins in $\vec{k}$
space was considered in the general case, and new
algorithm was suggested. The dynamics
in $\vec{k}$ space is 
no less easily captured
than in most standard methods that
work only in real space. 
No torques or forces need
be considered. The evolution 
of the spin in $\vec{k}$ space  
under the influence of the 
two spin interactions
is trivial.

\section{Appendix}

 Here we follow the beautiful treatment of  \cite{Nielsen}.
Within the (hard spin) fully constrained XY model:
\begin{eqnarray}
G(\vec{x}-\vec{y}) = \langle \vec{S}(\vec{x}) \cdot \vec{S}(\vec{y})
\rangle 
= \langle \cos[\theta(\vec{x})-\theta(\vec{y})] \rangle.
\end{eqnarray}
Here $\theta(\vec{x}) =  \vec{q}\cdot \vec{x} + \Delta \theta(\vec{x})$,
i.e. $\Delta \theta$ denotes the phase fluctuations about
our spiral ground state and 
\begin{eqnarray}
G(\vec{x}-\vec{y}) = \cos(\vec{q} \cdot (\vec{x}-\vec{y})) 
 \langle e^{i(\Delta \theta(\vec{x})- \Delta \theta(\vec{y})} \rangle.
\end{eqnarray}
In our harmonic approximation
$\{ \delta \theta(\vec{x})\}$
are random Gaussian variables
and only the first term in the 
cumulant expansion
is non-vanishing. The correlator 
\begin{eqnarray}
G(\vec{x})= \exp [-\frac{1}{2}[ \langle |\Delta \theta(\vec{x})-\Delta 
\theta(0)|^{2}] \rangle] \nonumber
\\ =  \exp 
\left[ 
k_{B}T 
\int 
\frac{d^{d}k}
{(2 \pi)^{d}}~ 
~\frac{1-\cos \vec{q} \cdot \vec{x}}
{A_{||} \delta_{||}^{2}+ A_{\perp}
\delta_{\perp}^{4}} 
\right].
\end{eqnarray}
Now let us shift variables $\vec{k} \rightarrow  
\vec{k} - \vec{q} \equiv \delta$, and for purposes
of convergence explicitly introduce an upper 
bound on 
$k_{\perp}$: ~$0<k_{\perp}< \Lambda$ 
\begin{eqnarray} 
I(\vec{x}_{\perp},x_{||}) \equiv \int \frac{1- \cos(\vec{q} \cdot \vec{x})}
{A_{||} k_{||}^{2}+ A_{\perp} k_{\perp}^{4}}.
\label{IIint}
\end{eqnarray}
This may be computed by first integrating 
over $k_{||}$ employing
\begin{eqnarray}
\int_{-\infty}^{\infty} \frac{1-\cos[a(b-x)]}{x^{2}+c^{2}} dx = \frac{\pi}{c}[1-e^{-ac} \cos(ab)],
\end{eqnarray}
to obtain
\begin{eqnarray} 
\frac{1}{A_{||}}
\int_{0}^{\infty} \frac{1-\cos(k_{||}x_{||} +
 \vec{k}_{\perp} \cdot \vec{x}_{\perp})}
{k_{||}^{2}+ (A_{\perp} k_{\perp}^{4}/A_{||})} dk_{||} \nonumber
\\ = \frac{1}{2} \frac{\pi}{k_{\perp}^{2}}
\sqrt{\frac{1}{A_{||} A_{\perp}}}
[1- \exp(-\sqrt{\frac{A_{\perp}}{A_{||}}} \vec{k}_{\perp}^{2} x_{||})
\cos (\vec{k}_{\perp} \cdot \vec{x}_{||})].
\end{eqnarray}
If $\phi$ denotes the angle between $\vec{k}_{\perp}$ and 
$\vec{x}_{\perp}$ then
\begin{eqnarray}
\int_{0}^{2 \pi} 
[1-\exp(-\sqrt{\frac{A_{\perp}}{A_{||}}} \vec{k}_{\perp}^{2}x_{||}) 
\cos(\vec{k}_{\perp} \cdot \vec{x}_{\perp})] d \phi \nonumber
\\ = 2 \pi - \exp(-\sqrt{\frac{A_{\perp}}{A_{||}}} k_{\perp}^{2} x_{||})
\int_{0}^{2 \pi} \cos(k_{\perp} x_{\perp} \cos \phi) d \phi.
\end{eqnarray} 
As 
\begin{eqnarray}
J_{0}(x) = \frac{1}{\pi} \int_{0}^{\pi} \cos(x \cos \phi) 
d \phi,
\end{eqnarray}
the integral of Eq.(\ref{IIint}),
\begin{eqnarray}
I(\vec{x}) = \frac{1}{2A_{||}} \pi 2 \pi \int_{0}^{\Lambda} 
\frac{1-\exp(-\sqrt{\frac{A_{\perp}}{A_{||}}} k_{\perp}^{2} x_{||})
J_{0}(k_{\perp} x_{\perp})}{\sqrt{\frac{A_{\perp}}{A_{||}}}
k_{\perp}^{2}} k_{\perp} dk_{\perp}.
\nonumber
\end{eqnarray}
We may now insert the series expansion of $J_{0}(x)$ and integrate
term by term. The Bessel function
\begin{eqnarray}
J_{0}(z) = \sum_{n=0}^{\infty} \frac{(-)^{n}z^{2n}}{2^{2n} (n!)^{2}}.
\end{eqnarray}
Comparing to the series for the exponential integral
\begin{eqnarray}
E_{1}(z) = -\gamma - \ln z - \sum_{n=1}^{\infty} (-)^{n} \frac{z^{n}}{n(n!)}
\end{eqnarray}
($\gamma$ is Euler's constant), we find
that 
\begin{eqnarray}
G(\vec{x}) \sim \frac{4 d^{2}}{x_{\perp}^{2}}~ \exp[- 2 \eta \gamma -
\eta E_{1}(\frac{x_{\perp}^{2} q}{4 x_{||}
\sqrt{\frac{A_{\perp}}{A_{||}}}})]~\times ~ \cos[qx_{||}],
\end{eqnarray}
where $\eta= \frac{k_{B}T}{8 \pi} \sqrt{\frac{A_{||}}{A_{\perp}}} 
=\frac{k_{B}T}{16 \pi}q$ (in the last equality Eq.(\ref{AAp})
is invoked), and 
$d= \frac{2 \pi}{\Lambda}$, 
with $\Lambda$ the 
ultra violet momentum cutoff.

\section{acknowledgments}

The bulk of this work
was covered in my thesis \cite{thesis}
in 1999. I wish to acknowledge
my mentors S. A Kivelson 
and J. Rudnick for allowing
students to pursue their 
own ideas. I am indebted
to many valuable conversations
with L. Chayes, S. A. Kivelson,
J. Rudnick, and P. G. Wolynes.

$^{*}$ Present address: Theoretical Division, Los Alamos National Laboratory, 
Los Alamos, NM 87545


\begin{thebibliography}{10}
\bibitem{2dCoulomb} We define 
the Coulomb kernel as that solving the Poisson equation
in $d$ dimensions. With this definition in hand, the 2-d Coulomb kernel 
depends logarithmically on the separation between the two 
particles (or spins).  Physical `two dimensional systems' can be of
many different types, e.g.: (a) charges confined to a single 2-d
sheet, which interact via 3-d (1/r) Coulomb interactions in empty space;
(b) the same, but with the sheet embedded in some surrounding dielectric
or conducting medium which modifies the Coulomb interaction (adding
screening); (c) a stack of 2-d sheets, where the carriers are confined
to a single sheet but can interact Coulombically with carriers on all
sheets; (d) an array of infinitely long wires, with uniform charging
along the wires, (e) the interaction
between magnetic vortices and superficially (f) the interaction between 
dislocation and disclination lines (or points
in this two dimensional context) in a
solid which, amongst other things, carry 
an additional direction dependent tensorial 
character. Type (a) might be a model for the quantum Hall effect;
(b) for a quantum Hall effect with conducting gate; (c) for holes in a
cuprate superconductor; and (d) might be applied to a stripe array,
looked at sideways, if the stripes are all long and straight. 
The Coulomb interactions that we are addressing 
are of the form of systems (d), (e), and (f).
\bibitem{explain_cubic} By ``cubic rotational symmetry'',
we allude to the symmetry transformations $\vec{k} \to \vec{k}^{\prime}$
that leave the hyper-cubic lattice Laplacian $\Delta(\vec{k})$ 
invariant. In the continuum limit $|\vec{k}| \ll 1$,
the Lattice Laplacian  $\Delta(\vec{k}) \to k^{2}$
and the ``cubic rotational symmetry''
amounts to the standard rotational symmetry,
i.e. the set of transformation
that conserve $ \vec{k}^{2} = (\vec{k}^{\prime})^{2}$.
\bibitem{one_third}
It is sufficient to consider a ``stripe'' state of period $\ell$ (even) in a
linear spin chain with the coupling, say, given by Eq. (\ref{Q}) in which
for simplicity only the first two terms are kept. 
If $S_i = +1$ for $i = 0, 1, \ldots, {\ell \over 2} -1$, and
$S_i = -1$ for $i = {\ell \over 2}, {\ell \over 2} +1, \ldots, \ell-1$, then
after a simple calculation we get $|S_k|^2 = 8 / ( 1 - \cos k)$ for
the square of the Fourier-transform of $S_i$ with $k = 2 \pi j /\ell$ and
$j$ odd, and 0 for $j$ even. The energy of one period for such a
configuration consists of two parts: the part due to the short-range
ferromagnetic coupling, computed directly to be equal to 4, and the
part due to the long-range interaction, which is ${Q \over 2 \ell} \sum_k
\Delta^{-1} (k)|S_k|^2$. Assuming that $\ell \gg 1$, we can expand the
cosine in the Taylor series and extend the summation in $k$ to
infinity. Then, the energy per spin becomes
$$
{E \over \ell} = {4 \over N} + {Q \ell^2 \over \pi^4} \sum_{n = 0}^\infty {1
\over (2 n + 1)^4} = {4 \over \ell} + {Q \ell^2 \over 96},
$$
where we employ $\sum_{n = 0}^\infty {1 \over (2 n + 1)^4} =
{\pi^4 \over 96}$. Minimizing this expression with respect to $\ell$ we
obtain the optimal period of the stripe in a viable
ground state,
$$
\ell_{viable_gs} = 4 \sqrt[3]{3}\, Q^{-1/3}.$$ Note
that this value does not match that 
suggested by dimensional analysis 
(the length $q^{-1}= Q^{-1/4}$). For $O(n \ge 2)$ spins,
the ground state $\vec{q}_{i}$
always exactly match modes minimizing the interaction
kernel $v(\vec{k})$. In Ising
systems, however, this is not the case.
\bibitem{intracate_comm}
To concretely highlight the subtleties in 
various frustrated incommensurate systems,
note that if, for example,  $Q=16$ in 
the Coulomb Frustrated Ferromagnet
(Eq.(\ref{Q})), $\vec{q} = (\pi,0,0) \in M_{Q}$, and 
$2 \vec{q} \equiv 0 (mod ~2 \pi)$. The fluctuation
matrix is diagonal in the $\vec{k}$ basis,
and the fluctuations are divergent at finite 
temperatures. An identical situation occurs for 
$\vec{q} = (\pi,\pi,0) \in M_{Q=64}$.
When $Q= 4$, $\vec{q} = (\frac{\pi}{2},0,0) \in M_{Q} $,
and it is easy to show that determining the eigenvalue spectrum
degenerates into a problem in two parameters ($\Delta_{2}, k_{1}$)
where  $\Delta_{2} \equiv  2\sum_{l=2}^{3} (\frac{3}{2}-\cos k_{l})$ 
( s.t. $\Delta(\vec{k} + \vec{q}) = \Delta_{2} +
 2 \cos k_{1}, ~\Delta(\vec{k}) = \Delta_{2} -2 \cos k_{1}$~),
The fluctuation integral about the chosen ground state exhibits a
($d-2$) dimensional minimizing manifold (parameterized, in our case,
by ($\Delta_{2},k_{1}$)~). 
Note that, at higher order commensurabilities,
the dimensionality of the 
minimizing manifold is low. In fact,
the interaction will no longer be diagonalized in $\vec{k}-$space.
\bibitem{smart}
Note that even if we had not known the ground states
we could still prove that there is
no magnetization. 
All one would need to do is to 
apply an infinitesimal magnetic
field 
\begin{equation}
\vec{h}(\vec{x}) = h \vec{S}^{ground-state}(\vec{x})
\end{equation}
with as yet an unknown ground state $\vec{S}^{ground-state}(\vec{x})$ .
Replacing any appearance of the minimizing 
wave-vector $\vec{q}$ in the Schwarz inequality 
and the definition of $B(\vec{k})$ by the more general 
wave-vector $\vec{\ell}$ and setting
$h \to 0^{+}$ one would arrive at 
the conclusion that $m_{\vec{\ell}}=0$
for each mode $\vec{\ell}$. 
Consequently,
\begin{eqnarray}
\langle S_{n}(\vec{x}) \rangle = \frac{1}{N} \sum_{\vec{\ell}} e^{i
\vec{l} \cdot \vec{x}} m_{\ell} = 0
\end{eqnarray}


Stated alternatively, by Parseval's theorem,
\begin{eqnarray}
\sum_{\vec{x}} \langle \vec{S}(\vec{x}) \rangle^{2} = \frac{1}{N^{2}} 
\sum_{\vec{l}} \langle \vec{S}(\vec{l}) \rangle \langle 
\vec{S}(-\vec{l}) \rangle. 
\end{eqnarray}
Taking careful note of the system size ($N$) 
dependence in Eq.(\ref{main}), 
\begin{eqnarray}
\sum_{\vec{x}} \langle \vec{S}(\vec{x}) \rangle^{2} = {\cal{O}}
(\frac{N}{\ln N})
\end{eqnarray}
and thus the average value of $|\langle \vec{S}(\vec{x}) \rangle|$
at any given site $\vec{x}$ diminishes as
${\cal{O}}(\sqrt{1/\ln N})$
which vanishes in the thermodynamic ($ N \rightarrow \infty$)
limit.
\bibitem{precision}

More precisely, everything 
to be claimed holds for 
the more general class of functions
$v(\vec{k})$ for which 
\begin{eqnarray}
\max_{\vec{u} \in B.Z.} [v(\vec{k}+ \vec{u}) + v(\vec{k} - \vec{u}) - 2
v(\vec{u})] \le B_{1} k^{2}
\end{eqnarray}
is obeyed for 
all momenta $\vec{k}$
with some finite positive constant 
$B_{1}$.
\bibitem{explain}
Looking at the real space form of the constraining
term it is readily seen that $\langle \psi| H_{1} | \psi \rangle \ge 0$. 
If $|\delta S| \ll 1$ this implies that the quadratic term in $\delta
S(\vec{x})$ originating from $H_{1}$ is non-negative definite.
For $S_{i=3}^{g}(\vec{x}) =0$, this quadratic term in $\delta
S_{3}(\vec{x})$ vanishes. The lowest eigenvalue $\lambda_{\min} =2$
corresponds to this vanishing ${\cal{O}} (\delta S ^{2}) $
contribution from $H_{1}$. An analogous occurs 
in the fluctuation analysis about a poly-spiral state.
\bibitem{ }
For a vanishing lower cutoff $\epsilon$ on $||\vec{k}|-q|$,
the fluctuations of an $n-component$ spin ($d=3$) 
within the Coulomb Frustrated Ferromagnet about a 
helical ground state are given by
\begin{eqnarray}
\frac{ \langle (\Delta \vec{S})^{2} \rangle }{k_{B}T} = \frac{(n-2) 
\sqrt{Q}}{4 \pi^{2} \epsilon} - \frac{1}{16 \pi}Q^{1/4} \ln|{\epsilon}|.
\end{eqnarray}
More generally, in $d>2$~dimensions, the leading order infrared contribution
reads
\begin{eqnarray}
\frac{(n-2)~Q^{(d-1)/4}}{2^{d}~\pi^{d/2}~\epsilon~\Gamma(d/2)} 
- \frac{\epsilon^{d-3}Q^{1/4}}{2^{d+1}~\pi^{(d-1)/2}~(d-3)~\Gamma(\frac{d-1}{2})}.
\end{eqnarray}
\bibitem{wigner}
Any given finite periodic spin
configuration (including 
any Wigner crystal), contains 
only a finite set of wave-numbers $\{\vec{q}_{i}\}$.
An interaction kernel $v(\vec{k})$ may always
be fashioned such that its minima occur
at $\vec{k} = \vec{q}_{i}$.

For those with interest in
real charged systems, we note that
as in the standard lattice gas 
mapping, any charge only
problem may be translated
onto a scalar spin problem
via the generalized $S_{i}=c(p_{i}-\overline{p})$
with $S_{i}$ the spin at site $i$, c a fixed 
constant (conventionally c= 2
for Ising systems),
$p_{i}$ the charge at the
given site, and 
$\overline{p}$ the average charge
density over the entire system
(the background charge
taken to be a half in
the standard Ising lattice
gas problem).
Within the spherical model,
only global normalization
of the spins is required
and the charge density
may be arbitrary at any given
site. As the Hamiltonian
is quadratic in $\{S_{i}\}$,
the average $\langle S \rangle = 0$,
self-consistently
implying that the average
charge density $\langle p \rangle = \overline{p}$.
\end{thebibliography}
\end{document}